\documentclass[amsmath,amssymb,preprint,showpacs]{revtex4}
\usepackage{graphicx}
\begin{document}

\title{Lyapunov modes in soft-disk fluids}
\author{Christina Forster}
\email{tina@ap.univie.ac.at}
\affiliation{Institut f\"ur Experimentalphysik, Universit\"at Wien,
 Boltzmanngasse 5, A-1090 Wien, Austria}
\author{Harald A. Posch}
\email{posch@ls.exp.univie.ac.at}
\affiliation{Institut f\"ur Experimentalphysik, Universit\"at Wien,
 Boltzmanngasse 5, A-1090 Wien, Austria}
\date{\today}
\begin{abstract}
 Lyapunov modes are periodic spatial perturbations of phase-space
  states of many-particle systems, which are  associated with the small positive or negative
  Lyapunov exponents. Although familiar for hard-particle systems in one, two, and
  three dimensions, they have been difficult to find for soft-particles.
  We present computer simulations for soft-disk systems in two dimensions and
  demonstrate the existence of the modes, where also Fourier-transformation methods
  are employed. We discuss some of their properties in comparison with equivalent hard-disk results.
  The whole range of densities corresponding to fluids is considered.  We show that it is not
  possible to represent the modes by a two-dimensional vector field of the
  position perturbations alone (as is the case for hard disks), but that the momentum perturbations
  are simultaneously required for their characterization. 
\end{abstract}

\maketitle

\section{Introduction}
For the last 50 years molecular dynamics simulations have decisively 
contributed to our understanding of the structure and dynamics of simple
fluids and solids~\cite{Alder:1959}. More recently, also the concepts of 
dynamical systems theory have been applied to study the
tangent-space dynamics of such systems~\cite{posch:1988a}. Of particular 
interest is the extreme sensitivity of the phase-space evolution to small 
perturbations. On average, such perturbations  grow, or shrink, exponentially 
with time, which may be  characterized by a set of rate constants, the 
Lyapunov exponents. The whole set of exponents is referred to as the Lyapunov 
spectrum. This instability is at the heart of the ergodic and mixing
properties of a
fluid and offers a new tool for the study of the microscopic dynamics. 
In particular, it was recognized very early that there is a close connection 
with the classical transport properties of systems in nonequilibrium stationary states 
\cite{hhp87}. For fluids in thermodynamic equilibrium, an analysis of the 
Lyapunov instability  is expected to provide an unbiased expansion of the 
dynamics into events, which, in favorable cases, may be associated with 
qualitatively different degrees of freedom, such as the  translation and 
rotation of linear molecules \cite{mph98}, or with the intra-molecular rotation 
around specific chemical bonds \cite{p05}.  

Since the pioneering work of Bernal~\cite{Bernal:1968} with steel balls, hard 
disks have been  considered the simplest model for a ``real'' fluid. With respect to 
the structure, they serve as a reference system for highly-successful 
perturbation theories of liquids~\cite{Hansen:1991,Reed:1973}.
Recently we studied the Lyapunov instability of such a model and found
\cite{mathII,pf04,hpfdz02,Milanovic:2002,EFPZ04} 
\begin{enumerate}
\item that the slowly-growing and decaying perturbations associated with
the non-vanishing Lyapunov  
exponents closest to zero may be represented as periodic vector fields
coherently spread out over the  
physical  space and with well-defined wave vectors $k$. Because of their similarity with
the classical modes of fluctuating hydrodynamics we refer to them as
Lyapunov modes. Depending on the boundary conditions, the respective
exponents are degenerate, 
and the spectrum has a step-like appearance in that regime.
\item that the fast-growing or decaying perturbations are localized in
space, and the number of particles actively contributing at any
instant of time vanishes in the thermodynamic limit.  
\end{enumerate}

Experimentally, the Lyapunov modes were found for hard-particle systems in one, two and 
three dimensions and for various boundary conditions, provided that  the system size, $L$,
in at least one direction is large enough for the discrete particles to
generate a recognizable 
wave-like pattern with a  wave number $k$. Their theoretical
understanding is based on the spontaneous breaking of the translational symmetry of
the zero modes $(k=0)$, which are associated with the vanishing Lyapunov
exponents and are a consequence of the conservation laws in the system 
\cite{eg00,MM01,MM04,WB04,EFPZ04,Morriss:2004}. 

Based on this experimental and theoretical evidence for hard-particle systems, it is natural 
to expect that the Lyapunov modes are a general feature of many-body systems with short-range 
interactions and that the details of the  pair potential should not
matter. However, already our very first simulations with soft disks
in two dimensions  \cite{hpfdz02,Dellago:2002,pf04} revealed a much
more complicated scenario.  Whereas the spatial localization of the fast-growing or decaying perturbations could  be easily verified, the mode structure for the slowly-growing or
decaying perturbations was elusive and could, at first, not be unambiguously detected. Here we
demonstrate that the modes for  soft sphere fluids do indeed exist. However, sophisticated
Fourier-transformation techniques are  
needed  to prove their existence. Interestingly, the degeneracy of the  Lyapunov exponents
and, hence, the step structure of the spectrum so familiar from the hard-disk case is recovered,  
but only for low densities. This suggests that kinetic theory is a proper theoretical framework in that case. For intermediate and large-density soft-disk fluids the degeneracy disappears. We do not have an explanation for this fact. Most recently,  Lyapunov modes were
demonstrated by Radons and Yang \cite{ry04a,ry04b} for one-dimensional Lennard-Jones  
fluids at low temperatures and densities.    

In this work we analyze the Lyapunov instability of repulsive Weeks-Chandler-Anderson (WCA)
disks in two dimensions and compare it to analogous hard-disk
results. The main difficulty is the large box size required for the
modes to develop and, hence, the large number of particles, $N$. It requires 
parallel programming for the computation of the dynamics both in phase and tangent space and 
for the re-normalization of the perturbation vectors according to the classical algorithms of Benettin  
{\em et al.} \cite{Benettin} and Shimada  {\em et al.} \cite{Shimada}. In Sec. \ref{Num} we characterize  
the systems and summarize the numerical methods used. In Sec. \ref{lyap}  we discuss the surprising 
differences found between the Lyapunov spectra for the soft- and hard-particle fluids, particularly at 
intermediate and large densities. Sec.\ref{ch_van} is devoted to an analysis of the
zero modes associated with the vanishing Lyapunov exponents.
The Lyapunov modes are analyzed in Secs. \ref{ch_mode} and \ref{FT}. We close with some
concluding remarks in Sec. \ref{remarks}.

\section{Characterization of the system and methodology} 
\label{Num}
We consider  $N$ disks with equal mass $m$ in a two-dimen\-sio\-nal rectangular box with 
extensions $L_x$ and $L_y$ and aspect ration $ A \equiv
L_y/L_x$. Periodic boundary conditions 
are used throughout. The particles  interact with a smooth repulsive
potential, for which we consider two cases: \\
i) a  Weeks-Chandler-Anderson potential,
\begin{equation}
\phi_{WCA}=\left\{\begin{array}{ll}4\epsilon\left[\left(\frac{\sigma}{r}\right)^{12}-
                                       \left(\frac{\sigma}{r}\right)^6\right]+\epsilon,
                                       &\qquad  
                                       r \leq 2^{1/6}\sigma\\ 
                                      0,&\qquad r>2^{1/6}\sigma\end{array}\right. ,
\end{equation}
with a force cutoff at  $r = 2^{1/6} \sigma$. As usual, $\sigma$ and
$\epsilon$ are interaction-range 
and energy parameters, respectively, which are ultimately set to
unity. Actually, such a potential  is  
not the best choice, since the forces and their derivatives  (required
for the dynamics in tangent-space) are not continuous at the
cutoff. This introduces additional noise into the simulation and  
violates the conservation laws. It also affects the computation of the Lyapunov spectrum.
To avoid this problem we sometimes also use \\
ii) a power-law potential,
\begin{equation}
\phi_{PL} = \left\{\begin{array}{ll}100 \epsilon \left[ 1-
       \left(\frac{r}{\sigma}\right)^2\right]^4,  
       &\qquad r \leq \sigma \\
            0,&\qquad r>\sigma\end{array}\right.
\label{pl}
\end{equation}
which looks very similar but does not suffer from this deficiency. We mention, however,
that the results discussed below turn out to be insensitive to this
weakness of the WCA potential. 
For reasons of comparison we also consider \\
iii) a hard-disk potential,
 \begin{equation}    
      \phi_{HD} = \left\{\begin{array}{ll} \infty, &\qquad r \leq \sigma \\
            0,&\qquad r>\sigma\end{array}\right.
\label{hd}
\end{equation} 
The phase space, $X$,   has $4N$ dimensions, and a 
phase point  $\Gamma \in X$ is given by the $4N$-dimensional vector
$\Gamma(t)=\{q_i,p_i; i=1,\dots, N\}$,  
where $q_i$ and $p_i$, denote the respective positions and
momenta of the disks. It evolves according to the  time-reversible motion equations
\begin{equation}
\dot\Gamma=F(\Gamma),
\label{eqm}
\end{equation} 
which are conveniently written as a system of first order differential equations. Here, 
$ F(\Gamma) = \{\frac{p_i}{m}, -\frac{\partial\Phi}
{\partial q_i}; i = 1,\dots,N\}$ follows 
from Hamilton's equations, where $\Phi \equiv \sum_i\sum_{j>i}
\phi(|q_i - q_j|) $ is the total potential energy.
 
Any infinitesimal perturbation  
$\delta \Gamma_l =  \{\delta q_i^{(l)}, \delta p_i^{(l)};  i=1,\dots, N\}$ lies in 
the tangent space $T X$, tangent to the manifold $X$ at the phase point
$\Gamma(t)$. Here, $\delta q_i^{(l)},$ and $\delta p_i^{(l)}$ are two-dimensional
vectors and denote the position and momentum perturbations contributed by particle $i$.
 $\delta \Gamma_l$ evolves according to the linearized equations of motion
\begin{equation}
\dot{\delta \Gamma_l} = \frac{\partial F}{\partial \Gamma_l}    
                  \cdot \delta \Gamma_l.
\label{leqm}
\end{equation}
Oseledec's  theorem  \cite{Oseledec:1968} assures us that there exist  $4N$ orthonormal
initial tangent vectors $\delta\Gamma_l(0), l = 1,\dots,4N$, whose norm grows or
shrinks exponentially with  
time such that the Lyapunov exponents
\begin{equation}
\lambda_l=\lim_{t \to\infty} \frac{1}{t}\ln\frac{||\delta
    \Gamma_l(t)||}{||\delta\Gamma_l(0)||},\quad 
l=1,\dots,4N ,
\end{equation}
exist. However, in the course of time these vectors would all get exponentially close to the  
most-unstable direction and diverge, since the unconstrained flow in tangent space does not preserve
the  orthogonality of these vectors. In the classical algorithms of Benettin {\em et  al.}
\cite{Benettin}, and  Shimada  {\em et  al.} \cite{Shimada} also used in this work, this difficulty is
circumvented by replacing these vectors by a set of modified tangent
vectors, ${\delta_l, l =1, \dots, 4N} $, which are periodically
re-orthonormalized with a Gram-Schmidt procedure.  The Lyapunov
exponents are determined from the time-averaged renormalization
factors. The vectors $\delta_l $ represent the perturbations associated with $\lambda_l$ and are the 
objects analyzed below. Henceforth we use the notation $\delta q_i^{(l)},$ and $\delta p_i^{(l)}$ 
also for the individual particle contributions to the {\em ortho-normalized} vectors, 
$\delta_l =  \{\delta q_i^{(l)},\delta p_i^{(l)}; i=1,\dots, N\}$. For later use, we 
introduce also the squared norm in the single-particle $\mu$ space,
\begin{equation}
\gamma_i^{(l)} \equiv  \left( \delta q_i^{(l)}\right)^2 + \left( \delta p_i^{(l)}\right)^2.
\label{gammai}
\end{equation}
This quantity is bounded, $0 \leq \gamma_i^{(l)} \leq 1$, and obeys 
the sum rule $\sum_{i=1}^N \gamma_i^{(l)} = 1$ for any $l$.  Since the equations 
(\ref{leqm}) are linear  in $\delta \Gamma_l$, the quantities $\gamma_i^{(l)}$  indicate, how the 
activity for perturbation growth, as measured by $\lambda_l$, is distributed over the particles at any
instant of time. 

For  the hard-disk systems the motion equations (\ref{eqm}) and (\ref{leqm}) need to be
generalized to include  also the collision maps due to the instantaneous particle
collisions. For algorithmic details we  refer to our previous work \cite{dph96,dp97,mathII}. 

The exponents are taken to be ordered by size, $\lambda_l \geq \lambda_{l+1}$, where the index $l$ 
numbers the exponents. According to the conjugate pairing rule for symplectic
systems~\cite{Evans:1990,Ruelle:1999}, the exponents appear  in pairs such that the 
respective pair sums vanish, $\lambda_l+\lambda_{4N+1-l}=0$. Only the positive half 
of the spectrum needs to be calculated. Furthermore, six of the exponents,  
$\{\lambda_{2N-2\leq l\leq2N+3}\}$, vanish as a consequence of the constraints imposed by
the conservation of energy, momentum, center of mass (for the tangent-space dynamics
only), and of the non-exponential time evolution of the perturbation vector in the direction of the flow.
 
In the soft-disk case we use reduced units for which the disk diameter $\sigma$, the particles' mass 
$m$, the potential parameter $\epsilon$, and Boltzmann's  constant $k_B$
are all set to unity.  As usual, the temperature is defined according to $\langle K \rangle=
\left\langle\sum \frac{p^2}{2m}\right\rangle= (N-1)k_BT$,where $K$ is the total kinetic energy 
and $\langle \dots \rangle$ is a time average. For  hard disks $K$ is a
constant of the motion,  and $K/N$ is taken as the unit of energy. Since Lyapunov exponents for
hard disks strictly scale with the particle velocities and, hence, the
square root of the temperature, all comparisons between soft and hard
disks below are for equal temperatures, $T = 1$. 
 
The computation of full Lyapunov spectra for soft particles is a numerical challenge, since the number 
of simultaneously integrated differential equations increases with the square of the particle number. 
Therefore, parallel processing with up to 6 nodes is used, both for the integration and for the 
Gram-Schmidt re-orthonormalization procedure \cite{lingen:2002}.

\section{Lyapunov spectra for soft-disk fluids}
\label{lyap}

\subsection{Phenomenology}

\begin{figure}
 \centering
     {\includegraphics[width=7cm,angle=-90]{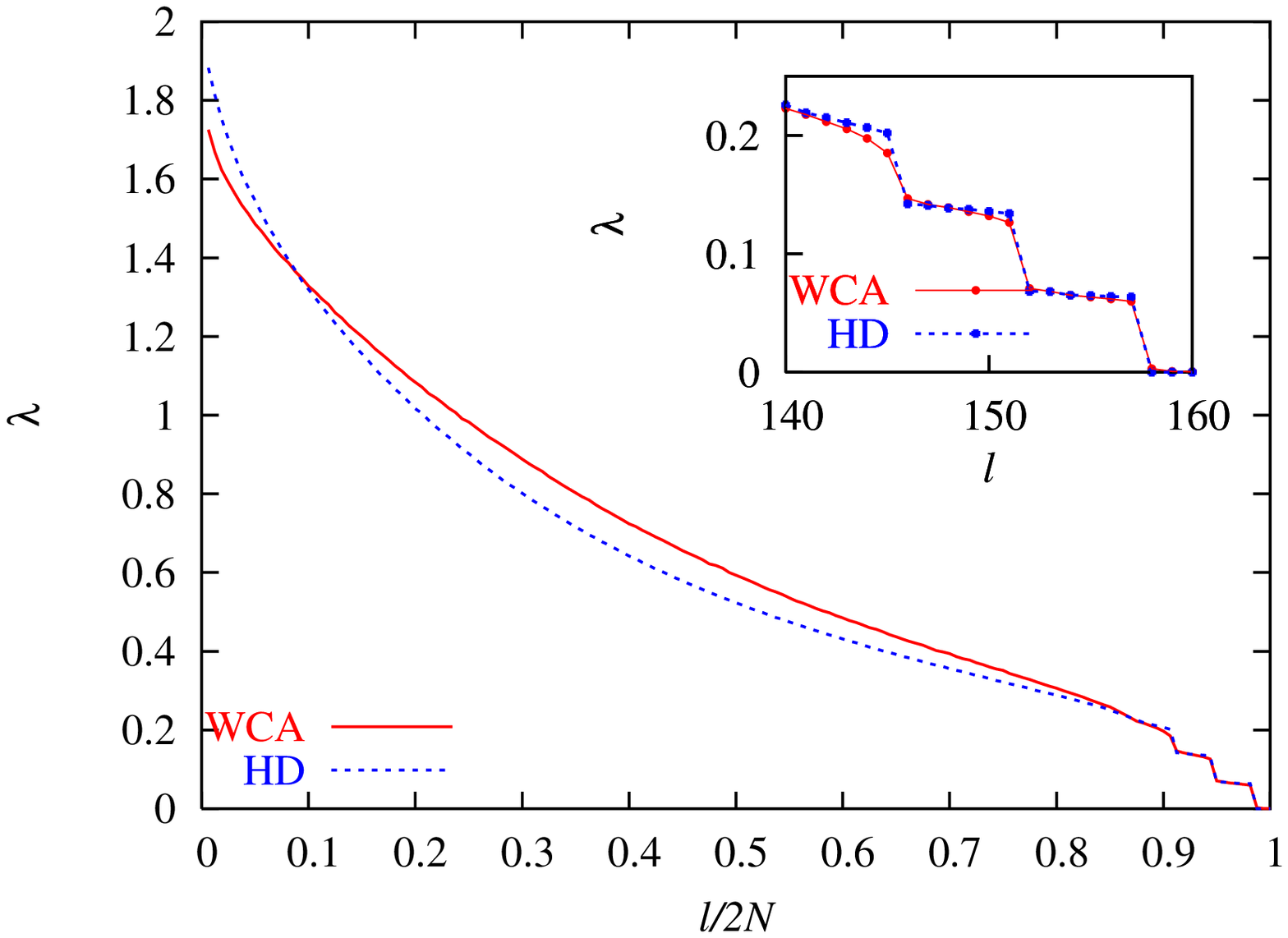} \\
       \includegraphics[width=7cm,angle=-90]{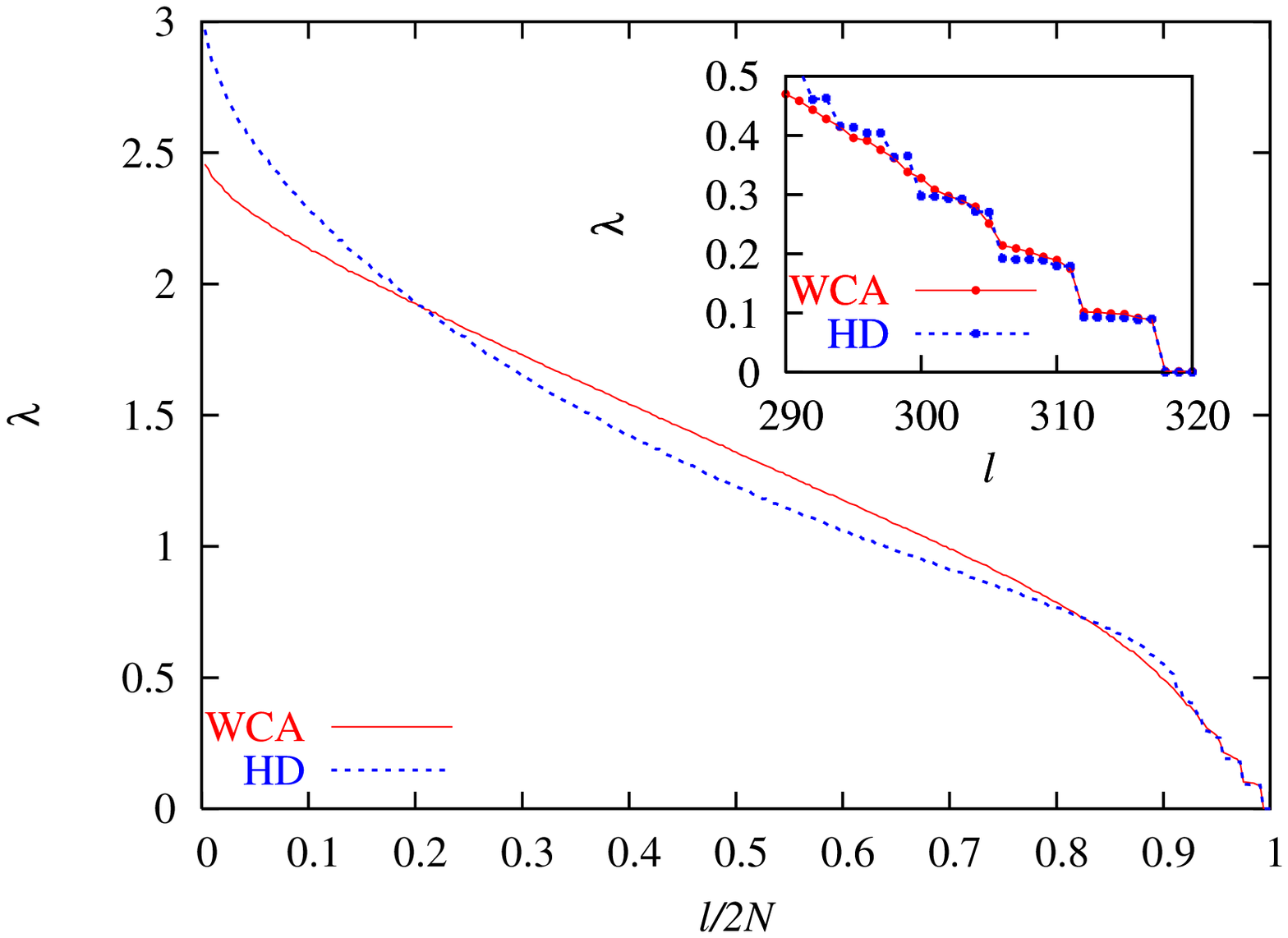}}
     \caption{Lyapunov spectra for soft (WCA) and hard (HD) disks at low 
       ($\rho=0.2$, upper panel)  and intermediate ($\rho=0.4$, lower panel) densities. The 
       very elongated rectangular simulation box is the {\em same} in both cases
       $(L_x = 100, L_y = 4, A = 0.04),$ such that $N=80$ (top) and $N=160$ (bottom)
       disks are involved, respectively. The temperature $T=1$. The Lyapunov index $l$  is normalized  
       by   $2N$. Only the positive branches of the spectra are shown. Although the spectra are only  
       defined  for integer values of $l$, smooth lines are drawn for clarity. In the inset a 
       magnified view of  
       the small-exponents regime is shown, where $l$ is not normalized.} 
\label{Fig_1} 
\end{figure}

We start our comparison of the spectra for soft (WCA) and hard-disks with low-density gases,   
$\rho \equiv N/L_x L_y =  0.2$, at a common temperature $T = 1$. 
The system consists of $N=80$ particles in a rather {\em elongated} rectangular simulation box,  
 $L_x = 100,  L_y = 4; A = L_y / L_x =0.04$, with periodic boundary conditions. An inspection 
of the upper panel of Fig.  \ref{Fig_1} shows, as expected, that the two systems have similar spectra 
and, thus, similar chaotic properties: the maximum exponents, $\lambda_1$, and the 
shapes of the spectra are reasonably close.  For even lower densities
(not shown) the agreement is even better. Most important, however, is
the observation that the step structure due to the degeneracy of the  
{\em small} Lyapunov exponents is observed both for the hard disks, for which it is well known to exist 
\cite{dph96,TM02,fhph04,EFPZ04},  and for the
soft-disk gases at low densities.  We conclude from this comparison in 
Fig. \ref{Fig_1} that the Lyapunov modes exist for low-density soft disks. In view of the smallness of
$L_y$ and the quasi one-dimensional nature of the system, these modes
may have only wave vectors parallel to the $x$ axis of the box. 

This general picture also persists for intermediate gas densities, $\rho= 0.4,$
as shown in the lower panel of Fig. \ref{Fig_1}. The box size is the same as before, but
the number of particles (and, as a consequence, also the number of exponents) is doubled. 
The spectral shapes for the  WCA and  hard-disk fluids, and the maximum exponents $\lambda_1$ 
in particular, become significantly different. Chaos, measured by  $\lambda_1$, is obviously 
enhanced for the WCA particles with a finite collision time  as compared to the instantaneously
colliding hard disks. Interestingly,  the Kolmogorov-Sinai (or dynamical) entropy $h_{KS}$, which 
is equal to the sum  of all positive exponents \cite{Pesin,ER85}, differs less due
to the compensating effect of the intermediate  exponents in the range $0.2 < l/2N < 0.8$. 
This may be verified in Fig. \ref{vergl_l1_hks} below.

\begin{figure}
 \centering
   {\includegraphics[width=6cm,angle=-90]{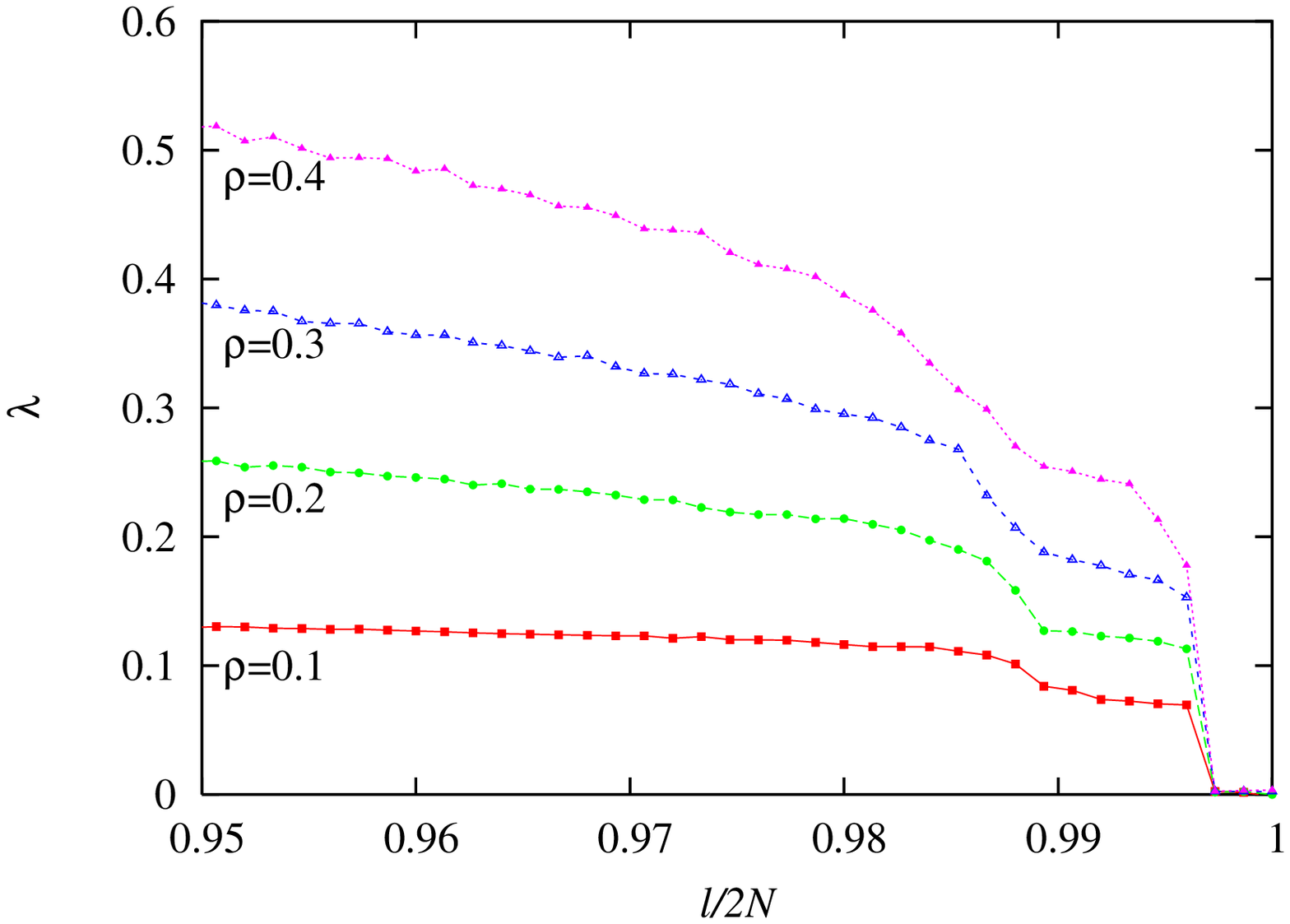}}
   \caption{Lyapunov spectra for WCA-disk fluids with $N=375$ particles in a periodic
    box with fixed aspect ratio $A = 0.6.$ Only the small exponents  are
    shown as a function of the reduced index $l/2N$. The density is indicated by  the labels.}
 \label{Fig_2}
\end{figure}

\begin{figure}
 \centering
   {\includegraphics[width=7cm,angle=-90]{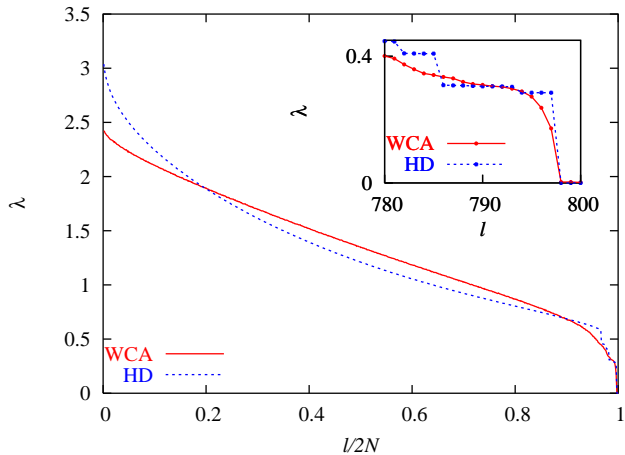}}
   \caption{Lyapunov spectrum for WCA and hard-disk fluids with $N=400$ particles,
    a density $\rho=0.4$, and $L_x=L_y=31.62$. The temperature $T=1$. The
    Lyapunov index $l$  is normalized  by   $2N$. The inset shows the lower part of the spectra
    where the step structure is prominent for the hard disks but totally  absent for the 
    WCA particles. For this part, the index is not normalized.}
 \label{Fig_3}
\end{figure}

The systems of Fig. \ref{Fig_1}  are quasi one-dimensional and do not represent bulk fluids. 
We show in Fig. \ref{Fig_2}  analogous spectra for {\em bulk} systems  in a rectangular simulation box
with aspect ratio $A = 0.6$, and containing $N=375$ particles. The density varies between 
0.1 and 0.4 and is  specified by the labels.
For all spectra the lowest step is clearly discernible, but is progressively less pronounced if 
the density is increased. For $\rho = 0.1$, the four smallest exponents  $(744 \leq l\leq 747)$
are identified as belonging to longitudinal modes (see below), the two next-larger
exponents $(742 \leq l\leq 743)$ to transverse modes.
In Figure \ref{Fig_3} we compare a WCA system to a hard-disk
system in a {\em square simulation box} at a density $\rho=0.4$ and $N=400$ particles. The
length of the simulation box in every direction is $L_x=L_y=31.62$. The
step structure that is so prominent in the hard disk spectrum vanishes altogether for
the soft potential case. If the density of such a square system with $N=400$ particles is
reduced by  increasing the size of the simulation box, the step structure with
the same degeneracy reappears.


\subsection{Measures for chaos}
\label{sec_measures}

\begin{figure}
  \centering{\includegraphics[width=6cm,angle=-90]{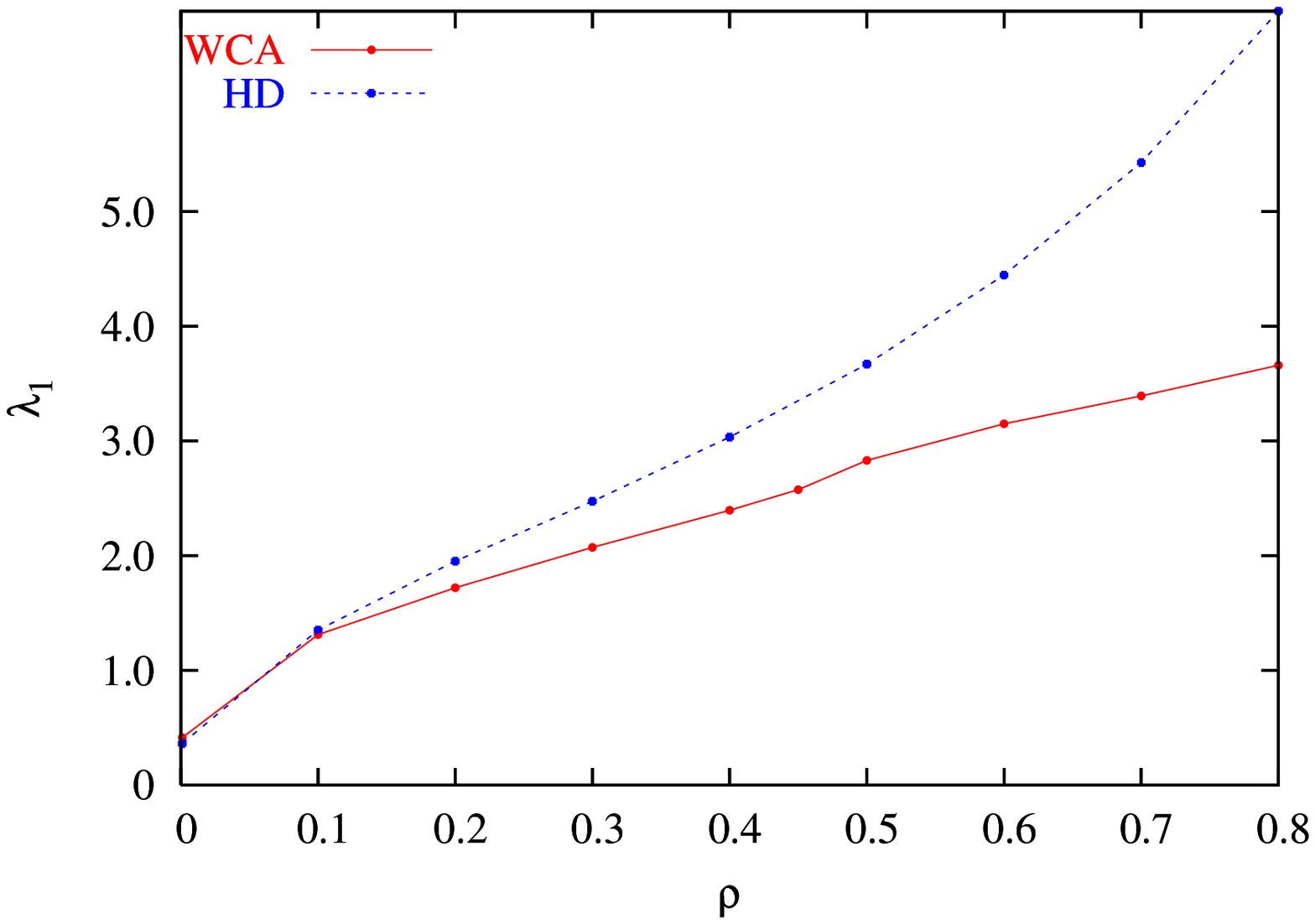}}
\vfill
    {\includegraphics[width=6cm,angle=-90]{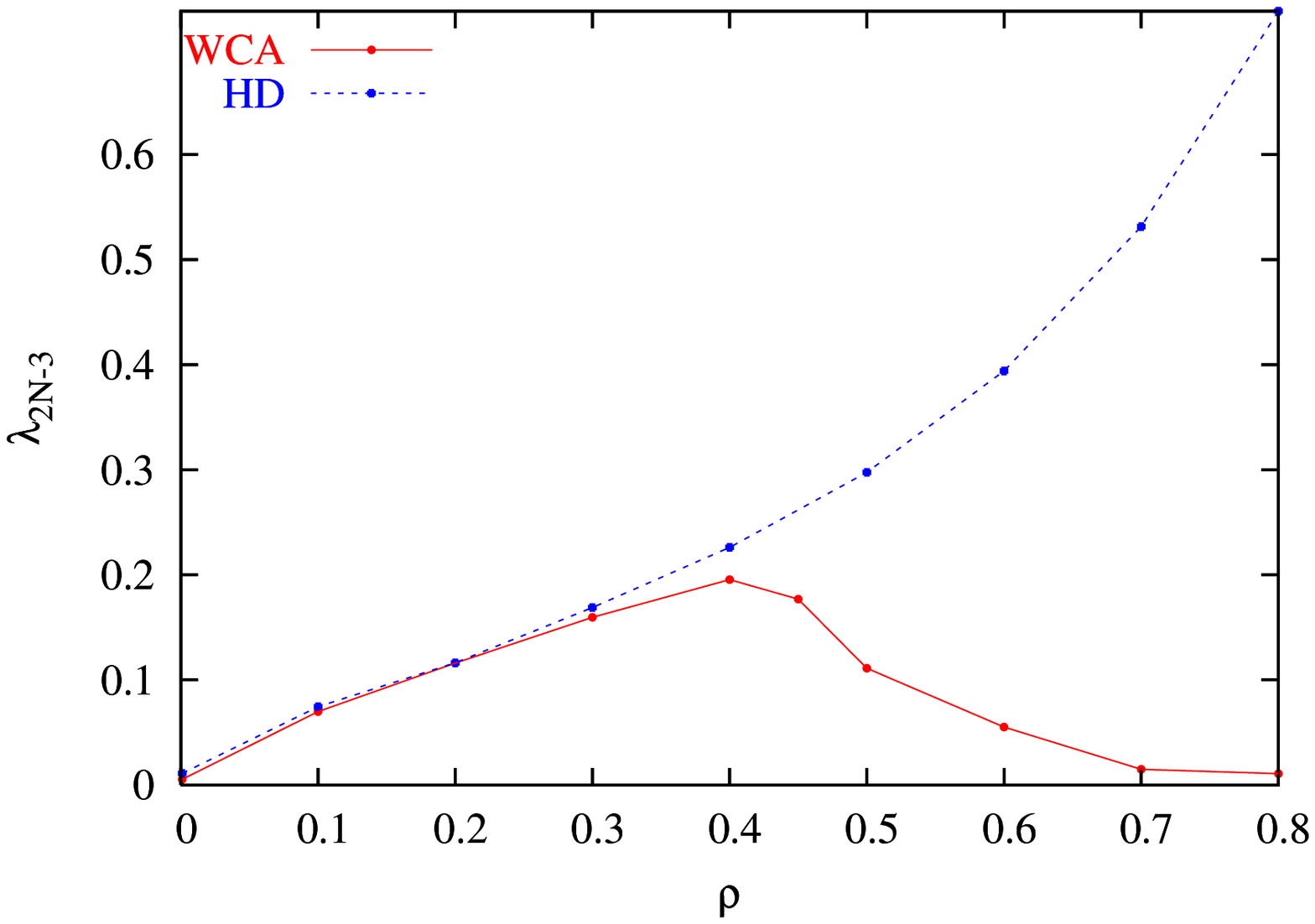}}
\vfill
    {\includegraphics[width=6cm,angle=-90]{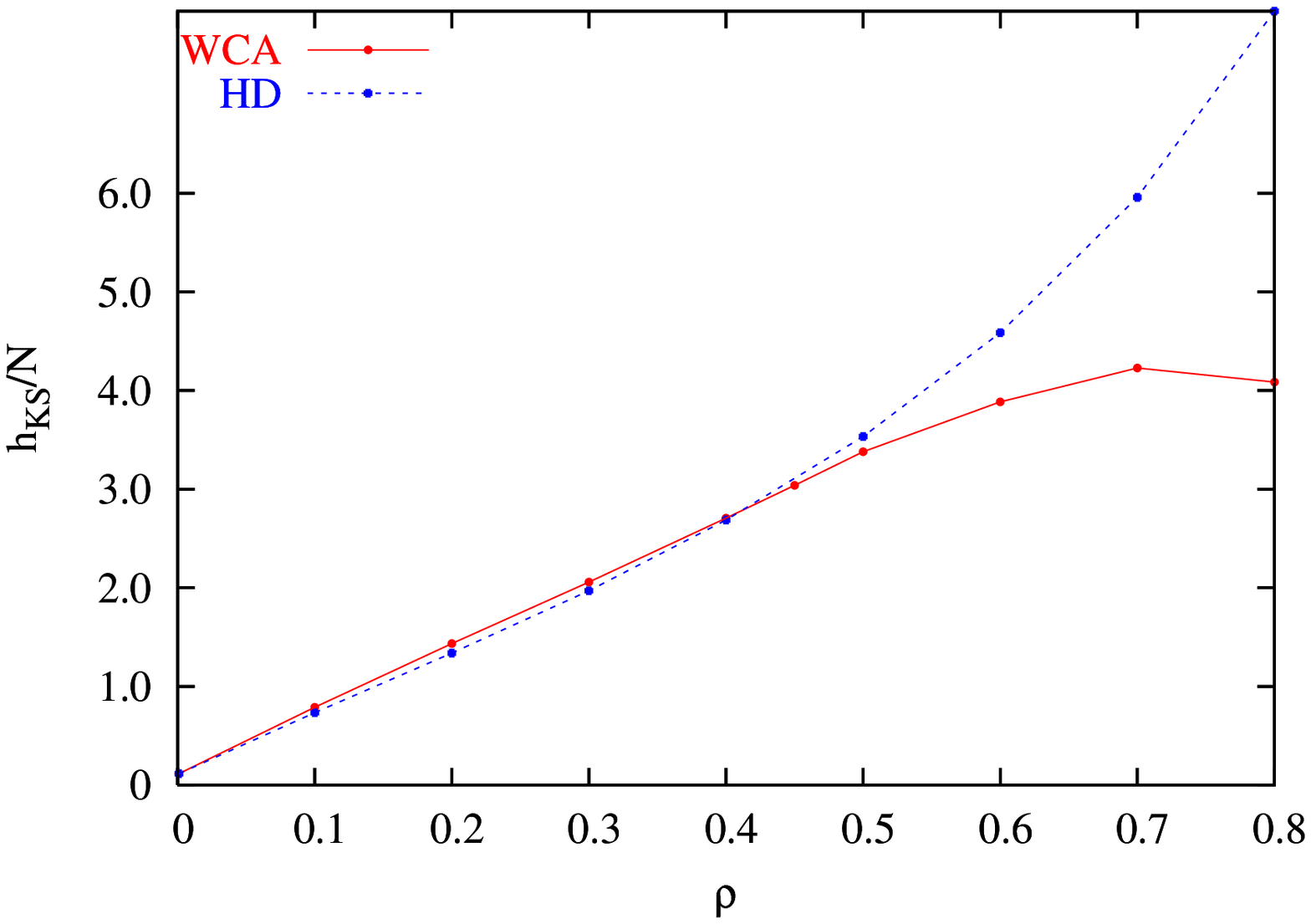}}
\vspace{3mm}
\caption{Isothermal density dependence of the maximum Lyapunov exponent, $\lambda_1$ (top), of 
the smallest positive exponent, $\lambda_{2N-3}$ (middle), and of the Kolmogorov-Sinai entropy 
per particle, $h_{KS}/N$ (bottom), for hard and  soft-disk systems as a function of the density. 
The particle number, $N=375$, the aspect ratio, $A=0.6$, of the periodic box, and the temperature, 
$T=1$ are held fixed.}
\label{vergl_l1_hks}
\end{figure}

    The maximum Lyapunov exponent, $\lambda_1$, and the Kolmo\-go\-rov-Sinai entropy,
$h_{KS}$, are generally accepted as measures for chaos. The latter corresponds to the
rate of information generated by the dynamics and, according to Pesin's theorem
\cite{Pesin,ER85} is equal to the sum of the positive Lyapunov exponents, 
$h_{KS} = \sum_{\lambda_l > 0} \lambda_l$. 
In Fig.  \ref{vergl_l1_hks} 
we compare the isothermal density dependences of these quantities for  375-disk WCA systems with
corresponding hard-disk results. Also included in this set of figures is a comparison for the
smallest positive exponent, $\lambda_{2N-3}$, which is directly affected by the 
mechanism generating the Lyapunov mode with the longest-possible wave length (if it exists).   

      For very dilute {\em hard-disk} gases, $\lambda_1$ and $h_{KS}/N$ have been estimated 
from kinetic theory  by Dorfman, van Beijeren and van Zon \cite{Szasz}, who have provided expansions
to leading orders in $\rho$, $\lambda_1 = A \rho ( -\ln \rho + B ) + \dots $, and 
$h_{KS}/N = \bar{A} \rho (-\ln \rho + \bar{B}) + \dots$, with explicit expressions for the constants
$A,B$ and $\bar{A},\bar{B}$. These predictions were very successfully confirmed by 
computer simulations \cite{BDPD,ZBD}.  Computer simulations by Dellago, Posch and Hoover
also provided hard-disk results over the whole range of densities \cite{dph96}. $\lambda_1$
and  $h_{KS}/N$ were found to increase mono\-ton\-ously with $\rho$ with the exception of a loop
in the density range between the freezing point of the fluid ( $\rho_f = 0.88 $ \cite{Stillinger}) 
and the melting point of the solid
($\rho_m=0.91$ \cite{Stillinger}). These loops disappear if $\lambda_1$
and  $h_{KS}/N$, are plotted as a function of the collision frequency $\nu$ instead of the density
\cite{dph96}. If $\rho$ approaches the close-packed density $\rho_0 = 1.1547 /\sigma^2$, both
 $\lambda_1$ and  $h_{KS}/N$ diverge as a consequence of the divergence of $\nu$. 
 
        The results for the WCA disks reported here are for $0.1 \leq \rho \leq 0.8$, which covers
almost the whole fluid range. Simulations for rarefied gases and for solids are currently under way
\cite{Forster05}. Not unexpectedly, we infer from the top panel of Fig. \ref{vergl_l1_hks} that 
$\lambda_1$-WCA  approaches $\lambda_1$-HD for small densities $\rho < 0.1$. However,
The density dependence for larger $\rho$ is qualitatively different. $\lambda_1$ approaches a
maximum near the density  corresponding to the fluid-to-solid phase transition, which confirms
our  previous results for Lennard-Jones fluids and solids \cite{Bunsen}. The collective dynamics
at such a transition causes maximum chaos in phase space. In the solid regime (not included in 
Fig. \ref{vergl_l1_hks}) $\lambda_1$ drops with increasing density, which is easily understandable.

        The smallest positive exponent $\lambda_{2N-3}$ for soft-disk gases follows the hard-disk
density behavior more closely than $\lambda_1$ for lower densities as inferred from the middle panel 
of Fig. \ref{vergl_l1_hks}. This means that the associated collective dynamics is  affected 
less by the details of the pair potential. Significant deviations start to occur for $\rho > 0.3$. 
For $h_{KS}/N$ the agreement between soft and hard disks extends even to a larger range of 
densities, as may be seen in the bottom panel of Fig. \ref{vergl_l1_hks}. However, this observation 
is deceptive and does not necessarily signal a similar dynamics. It is a consequence of significant 
cancellation due to the exponents intermediate between the maximum and the small exponents
of the spectra such as in the bottom panel of Fig. \ref{Fig_1}. 

\subsection{Localization of tangent-space perturbations in physical space}
\label{sec_loc}
One may interpret the {\em maximum (minimum) Lyapunov exponent} as the rate constant for the fastest
growth (decay) of a phase-space perturbation. Thus, it is dominated by the fastest dynamical events, 
binary collisions in the case of particle fluids. Therefore, it does not come as a surprise that the 
associated tangent vector has components which are strongly localized in physical space
\cite{mph98}. Similar observations for other spatially-extended systems have been made before
by various authors \cite{Manneville:1985,Livi:1989,Giacomelli:1991,Falcioni:1991}. 
With the help of a particular measure for the localization\cite{Milanovic:2002}, we could show
that for both hard and soft disk systems the localization persists even in the thermodynamic limit,
such that the fraction of tangent-vector components contributing to the generation of $\lambda_1$
at any instant of time converges to zero with $N \to \infty$ \cite{fhph04,pf04}. The localization
becomes gradually worse for larger Lyapunov indices $l>1$, until it ceases to exist and 
(almost) all particles collectively contribute to the perturbations associated with the
smallest Lyapunov exponents, for which coherent modes are known (or believed)
to exist  \cite{fhph04,pf04}.

Here, we adopt another entropy-based localization measure due to Taniguchi and Morriss 
\cite{Morriss:2003} which  was successfully applied to quasi-onedimensional hard-disk gases. 
Recalling the definition of the individual particle contributions 
$\gamma_i^{(l)}$ to  $\delta_l ^2 ( \equiv 1)$ 
in Eq. (\ref{gammai}), one may  introduce  an entropy-like quantity
\begin{equation}
S^{(l)}=-\sum_{i=1}^{N}{\langle \gamma_i^{(l)}(t)\ln \gamma_i^{(l)}(t)\rangle} \;,
\end{equation}
where $\langle \dots  \rangle$ denotes a time average. The number
\begin{equation}
     W^{(l)} \equiv \exp\left( S^{(l)} \right)
\end{equation}
may be taken as a measure for the spatial localization of the perturbation $\delta_l$
associated with $\lambda_l$, $1\leq l \leq 2N$. It is bounded according to 
$1 \leq W^{(l)} \leq N$. The lower bound, $1$, indicates the most-localized state
with only one particle contributing, and the upper bound $N$ signals uniform contributions
by all of the  particles $\{ \gamma_i^{(l)} = 1/N, i=1,\dots, N \}.$ We shall refer to the
set of normalized localization parameters $\{W^{(l)}/N, l=1,\dots,2N\}$ as the {\em localization
spectrum}.

\begin{figure}
\centering{\includegraphics[width=7cm,angle=-90] 
  {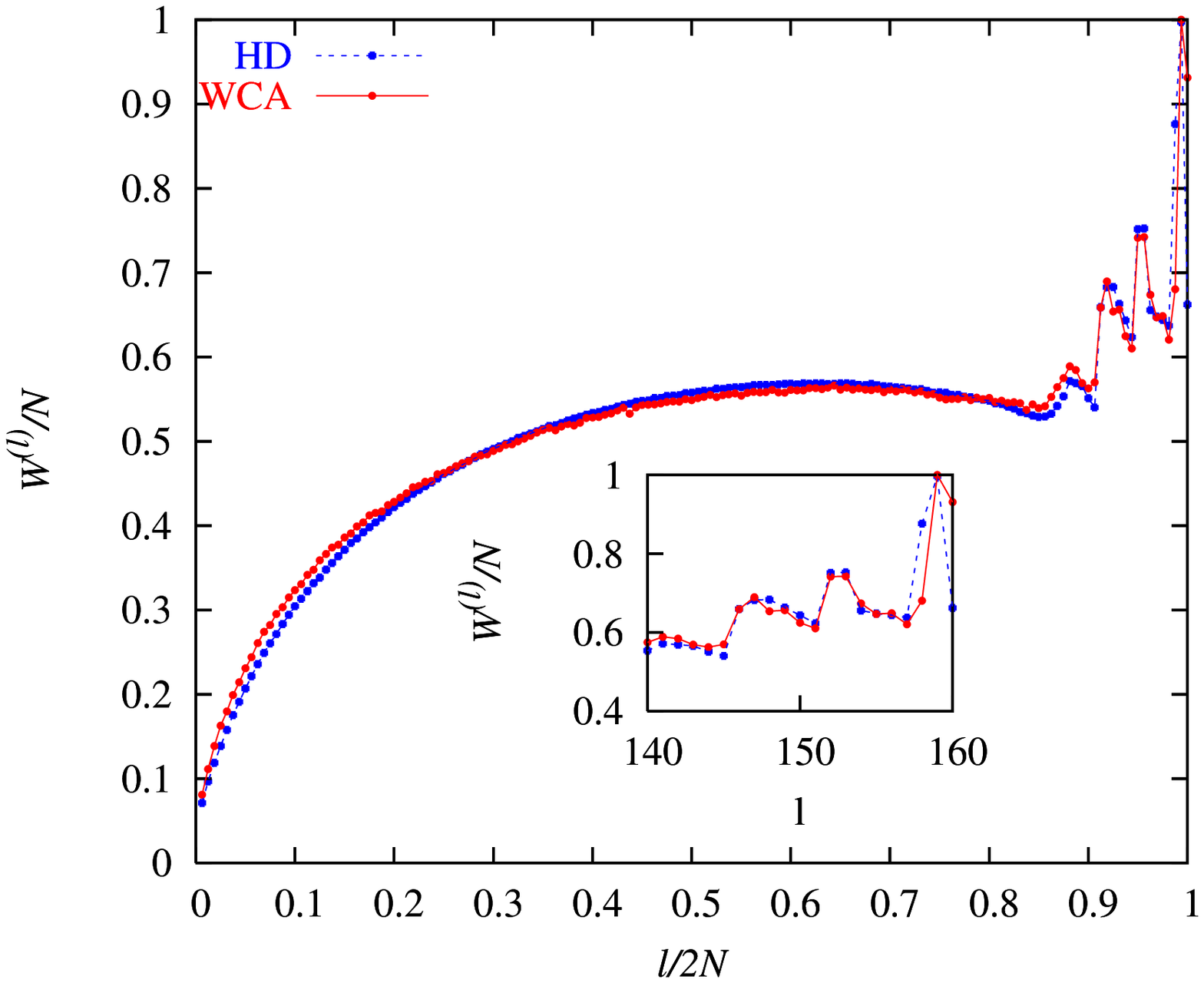}}\\
\centering{\includegraphics[width=7cm,angle=-90]{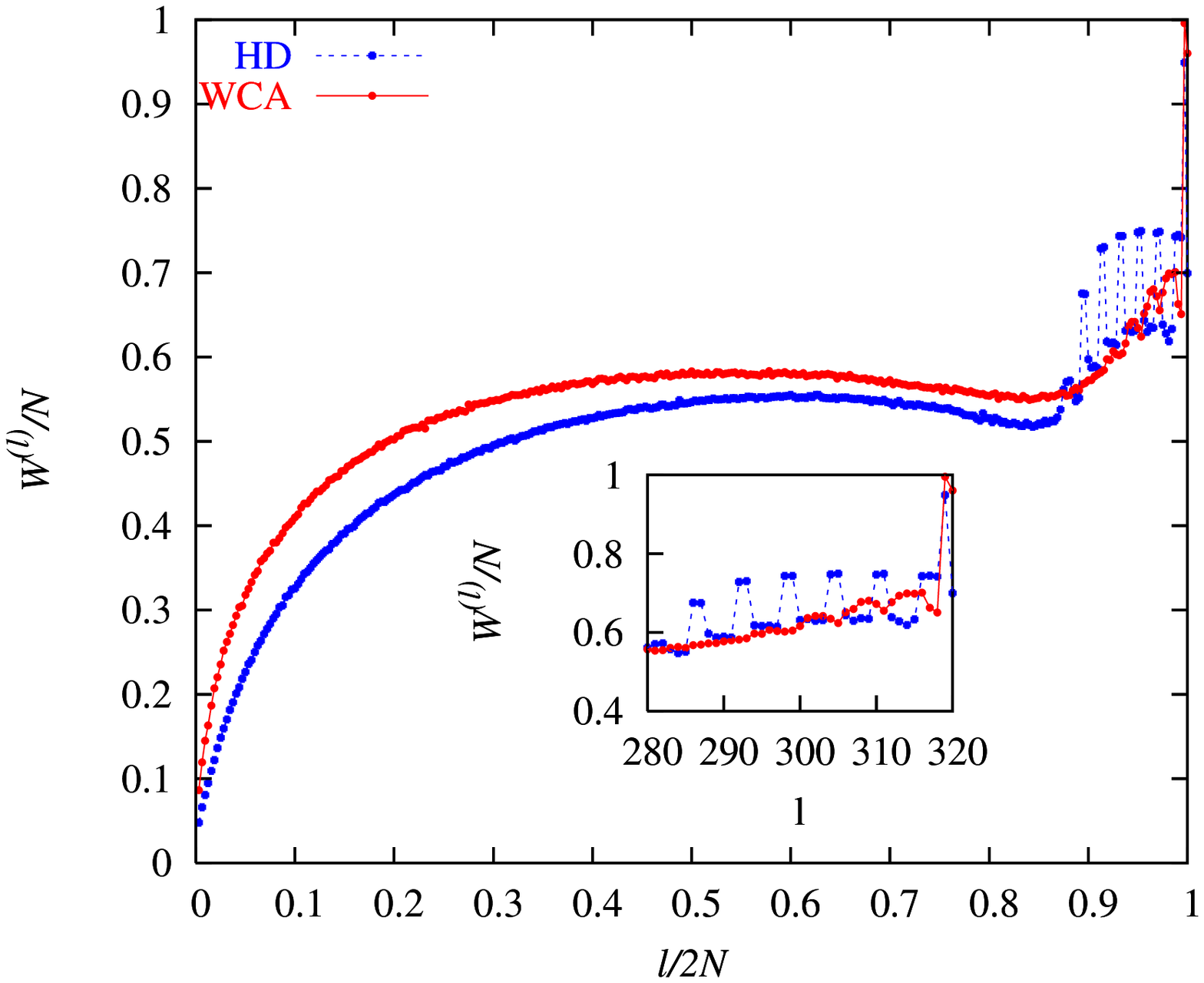}}
\vspace{3mm}
  \caption{Dependence of the normalized localization parameters,  $W^{(l)}/N$, on the
    normalized Lyapunov index $l/2N$. The corresponding Lyapunov spectra are
    shown in Fig. \ref{Fig_1}. The soft and hard disks are indicated by the labels and distinguished
     by  color. The system is a rectangular box ($L_x = 100$, $L_y = 4$) with
    periodic boundaries. Top: $\rho = 0.2$, $N = 80$. Bottom: $\rho = 0.4$, $N = 160$.
    The temperature $T=1$. The insets provide a magnified view of the mode regime.}
\label{loc_width}
\end{figure}
    In Fig. \ref{loc_width} we compare the localization spectra for the
same hard-disk and soft-disk systems for which the Lyapunov spectra are given in 
Fig. \ref{Fig_1} ($L_x = 100$, $L_y = 4$, $A = 0.04).$ For the lower-density gases in the top panel 
($\rho = 0.2$, $N = 80$), the localization spectra for soft and hard disks are almost indistinguishable. 
They indicate strong localization
for the perturbations belonging to the large exponents (small $l$), and collective behavior
for large $l$. Of particular interest is the comb-like structure also magnified in the inset,
which is a consequence of the Lyapunov modes (which will be discussed in more detail below). 
Stationary transverse modes are hardly affected by the 
time averaging involved in the computation of the localization measure and lead to large values 
of $W^{(l)}/N$, indicating strong collectivity. The propagating LP-modes \cite{EFPZ04}, however,
are characterized by a significantly-reduced value  of $W^{(l)}/N$. For
intermediate gas densities in the bottom panel ($\rho = 0.4$, $N = 160$), differences between
the hard and soft-disk results become apparent. Most prominently, the comb structure in 
the localization spectrum of the WCA system has mostly disappeared, although the Lyapunov
spectrum clearly displays steps in the small-exponent regime, which is a clear indication for the
existence of modes.


\section{Zero modes}
\label{ch_van}

The dynamics in phase space and tangent space is strongly affected by the inherent
symmetries of a system, i. e. infinitesimal transformations leaving the equations of motion invariant.
They are intimately connected with the conservation laws obeyed by the dynamics.
If these transformations act as infinitesimal perturbations of the initial conditions, the latter do 
not grow/shrink exponentially (but at most linearly)  with time and give rise to as many vanishing 
Lyapunov exponents. The generators of these transformations are taken as unit vectors in 
tangent space, which point into the direction of the respective perturbation and are called zero 
modes. They span an invariant subspace ${\cal N}(\Gamma)$ of the tangent space at any phase 
point $\Gamma$, 
which is referred to as the zero subspace. They are essential for an understanding of the Lyapunov 
modes dealt with in the following.

For a planar system of hard or soft disks, there are six symmetry-related perturbations leading to 
non-expo\-nen\-tial growth \cite{EFPZ04,WB04} and, hence, six vanishing Lyapunov exponents:
\begin{enumerate}
\item Homogeneous translation in $x$ and $y$ directions, with  generators $e_1$ and $e_2$ explicitly 
given below. It is a consequence of center-of-mass conservation, which is actually not obeyed
for systems with periodic boundary conditions, but, nevertheless,  still holds for the linearized motion in tangent   space, which does not recognize periodic box boundaries for the dynamics of  the 
perturbations.
\item  Homogeneous momentum perturbation in $x$ and $y$ directions due to Galilei invariance, 
with generators $e_3$ and $e_4$ given below. It may be viewed as a consequence of momentum conservation.
\item Simultaneous momentum and force rescaling according to a generator $e_5$ as
given below. It is a consequence of energy conservation and, ultimately, of the homogeneity of time.  
\item Homogeneous time shift with generator $e_6$, corresponding to non-exponential growth
in the direction of the phase flow and also a consequence of the homogeneity of time.
\end{enumerate}  
 
These generators follow from the constraints the conservation equations impose on the dynamics in 
phase space. The latter  constitute hyper-surfaces,   $\sum_{i=1}^N x_i=$ const, 
$\sum_{i=1}^N y_i$= const for the center of mass (see the comment above), 
$\sum_{i=1}^N p_{x,i} = 0$, $\sum_{i=1}^N p_{y,i} = 0$ for the conserved momentum,
and  $\sum_{i=1}^N (p_{x,i}^2 + p_{y,i}^2 )/2 +\Phi = E$ for the conserved energy. The
gradients perpendicular to these hyper-surfaces provide directions of
non-exponential phase-space growth and, hence, the generating vectors $e_1, \dots, e_5$
If we use the explicit notation
$$
\delta \Gamma = \{\delta x_i, \delta y_i, \delta p_{x,i}, \delta p_{y,i};i=1,\dots,N\}
$$
for an arbitrary tangent vector, one obtains
\begin{eqnarray}
e_1  & = &  (1/N) \{1,0,0,0 ;  i=1,\dots,N\} \label{e1}\\
e_2  & = &  (1/N) \{0,1,0,0 ;  i=1,\dots,N\} \label{e2}\\
e_3  & = &  (1/N) \{0,0,1,0 ;  i=1,\dots,N\} \label{e3}\\
e_4  & = &  (1/N) \{0,0,0,1 ;  i=1,\dots,N\} \label{e4}\\
e_5  & = &   \alpha \{-F_{x,i},-F_{y,i},p_{x,i},p_{y,i};  i=1,\dots,N\} \label{e5}\\
e_6  & = &   \alpha \{p_{x,i},p_{y,i}, ;F_{x,i},F_{y,i};  i=1,\dots,N\} ,\label{e6}
\end{eqnarray}
where $\alpha$ is a normalizing factor. Each vector has $4N$ dimensions. $e_1$  to $e_6$ are 
orthonormal and form a natural basis for the invariant zero space. In the following we
consider the expansion of the  six orthonormal tangent vectors
$\delta_{2N-2}, \dots ,\delta_{2N+3}$, responsible for the six vanishing exponents in the simulation,
in that basis,
\begin{equation}
       \delta_{2N-3+l}  = \sum_{j=1}^6 \alpha_{l,j} \; e_j \;, \;      l= 1, \dots, 6 \; ,
       \label{expansion}
\end{equation}
with projection cosines, $\alpha_{l,j} \equiv  (\delta_l \cdot e_j ) $. Since all vectors involved
are of unit length, $\alpha_{l,j}$ may either be interpreted as the projection of 
$\delta_{2N-3+l}$ onto $e_j$, or {\em vice versa}.

For {\em hard disk fluids} one can easily show \cite{EFPZ04} that the tangent vectors
$\delta_{2N-2}, \dots ,\delta_{2N+3}$ are fully
contained  in ${\cal N}( \Gamma)$.   The projection cosines strictly obey the sum rules
$$\sum_{i=1}^6 (\alpha_{l,i})^2 = 1\;\;\;, \;\;\;
  \sum_{l=1}^{6}  (\alpha_{l,i})^2 = 1 \; ,$$
at all times, if the total momentum vanishes and if $K = N$, as is always the case in our simulations. Thus,  ${\rm  Span}(\delta_{2N-2},\dots, \delta_{2N+3}) = {\rm  Span}(e_1,\dots, e_6).$ Note that there are
no forces acting on the particles in this case, and the particles are moving on straight lines between
successive instantaneous collisions.

\begin{figure}
\centering{\includegraphics[width=8cm] 
  {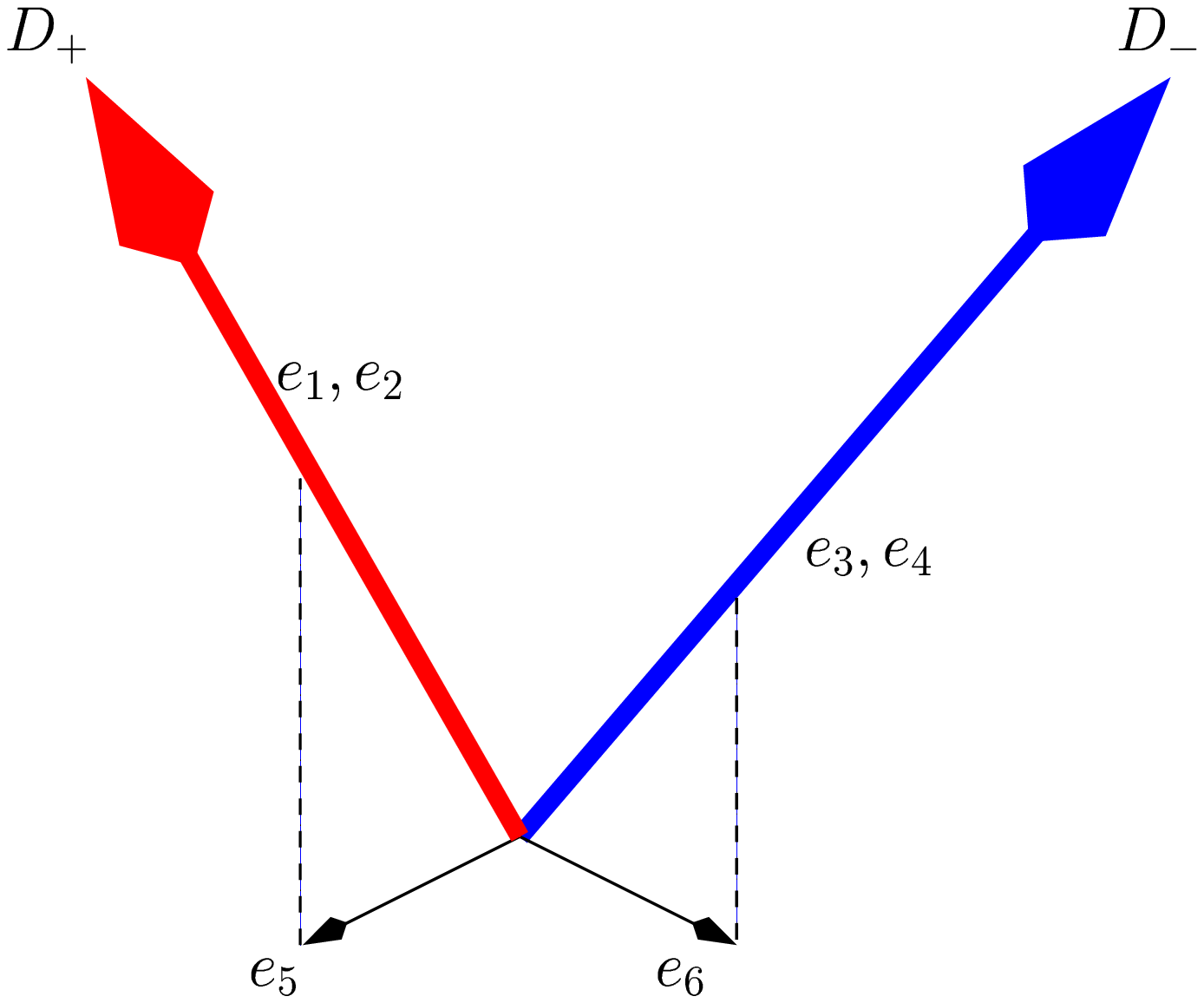}}\\
  \caption{Caricature of the high-dimensional tangent space for $N$ particles.  
  The orthogonal zero modes
  $e_1$ and $e_2$ always point into the subspace   ${\cal D}_{+} \equiv 
  {\rm  Span}(\delta_{2N-2},\delta_{2N-1}, \delta_{2N})$. Similarly, the zero modes  $e_3$ and $e_4$ always point into the subspace   ${\cal D}_{-} \equiv  {\rm  Span}(\delta_{2N+1},\delta_{2N+2}, \delta_{2N+3})$. The subspaces ${\cal D}_{+}$ and  ${\cal D}_{-}$ are represented by the thick arrows. The projection angle of the zero mode $e_5$ into ${\cal D}_{+}$ agrees with that of $e_6$ into
$ {\cal D}_{-}$ and generally differs from zero whenever a collision occurs anywhere
in the system.}
\label{proj_zero}
\end{figure}

For {\em soft disk systems}, however, the situation is more complicated.
\begin{table}
\caption{Projection cosines, $\alpha_{l,j}$, according to Eq. (\ref{expansion}) for $N=4$ soft particles
in a periodic box and interacting with the power-law  potential (\ref{pl}). The density is low, $\rho = 0.1$,
to isolate individual collision events. Two instants are considered:}
\label{cosines}
\begin{center}
        A: At time $t_0$ just before a collision of particles 1 and 2:\\
\begin{tabular}{c|rrrrrr|c} 
$  l $ & $\alpha_{l,1} $ & $\alpha_{l,2} $ & $\alpha_{l,3} $ & $\alpha_{l,4}
$ & $\alpha_{l,5} $ & $ \alpha_{l,6}  $&$\sum_{j=1}^{6}{(\alpha_{l,j})^2}$ \\ \hline
1 & -1.00000 & 0.00000 & 0.00000 & 0.00000 &  0.00000 &  0.00000 & 1.00000\\[-2mm]
2 &  0.00000 & 0.21120 & 0.00000 & 0.00000 & -0.00001 & -0.97734 & 0.99980\\[-2mm]
3 &  0.00000 & 0.97744 & 0.00000 & 0.00000 &  0.00000 &  0.21118 & 0.99999\\[-2mm]
4 &  0.00000 & 0.00000 & 0.95954 & 0.00000 &  0.28155 &  0.00000 & 0.99998\\[-2mm]
5 &  0.00000 & 0.00000 & 0.28158 & 0.00000 & -0.95944 &  0.00001 &${\it 0.99981}$\\[-2mm]
6 &  0.00000 & 0.00000 & 0.00000 & 1.00000 &  0.00000 &  0.00000 & 1.00000\\\hline
$\sum_{l=1}^6{(\alpha_{l,j})^2}$& 1.00000 & 1.00000 & 1.00000 & 1.00000 & ${\it 0.99979} $&
   $ {\it 0.99979} $ &\\[3mm]
\end{tabular}
       \\B: At time $t_0 + 0.11$ of closest approach of particles 1 and 2:\\ 
\begin{tabular}{c|rrrrrr|c}
$ l $ & $\alpha_{l,1} $ & $\alpha_{l,2} $ & $\alpha_{l,3} $ & $\alpha_{l,4}
 $ & $\alpha_{l,5} $ & $ \alpha_{l,6}  $&$\sum_{j=1}^{6}{(\alpha_{l,j})^2}$ \\ \hline
1 & -1.00000 & 0.00000 & 0.00000 & 0.00000 &  0.00000 &  0.00000 & 1.00000\\[-2mm]
2 &  0.00000 & 0.14132 & 0.00000 & 0.00000 &  0.00000 & -0.08601 & 0.02737\\[-2mm]
3 &  0.00000 & 0.98996 & 0.00000 & 0.00000 &  0.00000 &  0.01228 & 0.98018\\[-2mm]
4 &  0.00000 & 0.00000 & 0.98172 & 0.00000 &  0.01654 &  0.00000 & 0.96405\\[-2mm]
5 &  0.00000 & 0.00000 & 0.19033 & 0.00000 & -0.08529 &  0.00000 & ${\it 0.04350}$\\[-2mm]
6 &  0.00000 & 0.00000 & 0.00000 & 1.00000 &  0.00000 &  0.00000 & 1.00000\\\hline
$\sum_{l=1}^6{(\alpha_{l,j})^2}$& 1.00000 & 1.00000 & 1.00000 & 1.00000 &$ {\it 0.00755} $&
    $  {\it 0.00755}  $ &\\[2mm]
\end{tabular}
\end{center}
\end{table}
In Table \ref{cosines} we list instantaneous matrix elements $\alpha_{l,j}$ for a simple example,
$N=4$ particles interacting with the particularly-smooth repulsive potential of 
Eq. \ref{pl}. The density, $\rho =0.1$, is low enough such that isolated binary collisions may be easily
identified. Two instances are considered: one just at the beginning of a collision of  
particles 1 and 2 (upper part of the table), and a time half way through this collision (lower part of the 
table). Considering the columns first, the squared sums  $\sum_{l=1}^6{(\alpha_{l,j})^2}$
always add to unity for $j=1,\dots,4$, which means that $e_1$ to $e_4$ are fully contained in the subspace 
${\rm  Span}(\delta_{2N-2},\dots, \delta_{2N+3})$ of the tangent space. The same is not true,
however, for $e_5$ and $e_6$, which contain in their definition the instantaneous forces acting on the
colliding particles. If no collisions take place anywhere in the system (upper part),
the subspaces   ${\rm  Span}(\delta_{2N-2},\dots, \delta_{2N+3})$ and  ${\rm  Span}(e_1,\dots, e_6)$
are nearly the same, but not quite, as the squared projection-cosine sums indicate. The moment
a collision occurs, the sums $\sum_{l=1}^6{(\alpha_{l,j})^2}$ for $j=5$ and $6$ may almost vanish,
as happens in the bottom part of the table, and the vectors $e_5$ and $e_6$ become
nearly orthogonal to ${\rm  Span}(\delta_{2N-2},\dots, \delta_{2N+3})$. In Figure \ref{proj_zero}
an attempt is made to illustrate this relationship for such  a high-dimensional tangent
space. For systems containing many particles and/or for larger densities, there will always
be at least one collision in progress, and the horizontal sums, $\sum_{j=1}^6{(\alpha_{l,j})^2}$
for $l=1,\dots,6$, and the vertical sums,  $\sum_{l=1}^6{(\alpha_{l,j})^2}$ for $j=5$ and $6$,
fluctuate and assume values significantly smaller than unity.  The subspaces 
 ${\rm  Span}(\delta_{2N-2},\dots, \delta_{2N+3})$ = ${\cal D}_{+} \oplus {\cal D}_{-}$ (see the 
 definition in Fig. \ref{proj_zero}) and  ${\cal N} = {\rm  Span}(e_1,\dots, e_6)$
do not agree in general.


\section{Lyapunov modes}
\label{ch_mode}


Lyapunov modes are periodic spatial patterns observed for the perturbations
associated with the  small positive and negative Lyapunov exponents. They are
characterized by wave vectors $k$ with wave number
\begin{equation}
k_{n_x,n_y}=\sqrt{\left({\frac{2\pi}{L_x}}\,n_x\right)^2
+\left({\frac{2\pi}{L_y}}\,n_y\right)^2} \;;\;\; n_x, n_y=0,1,\ldots\;\;,
\label{kvectors}
\end{equation}
where a rectangular box with periodic boundaries is assumed, and  where $n_x$ and $n_y$ denote the
number of nodes parallel to $x$ and $y$, respectively. For modifications due to other 
boundary conditions we refer to Refs. \cite{EFPZ04,TM03}. Experimentally, Lyapunov modes were
observed for hard particle systems in one, two, and three dimensions 
\cite{mathII,hpfdz02,EFPZ04,Morriss:2003,Morriss:2004}, for hard planar 
dumbbells \cite{mph98,mph98a,Milanovic:2002}, and, most recently, 
for one-dimensional soft particles \cite{ry04a,ry04b}.
Theoretically, they are interpreted as periodic modulations $(k\neq0)$ of the zero modes, to which
they converge for  $k \to 0$.  Since a modulation, for $k > 0$, involves the  breaking of a 
continuous symmetry (translational symmetry of the zero modes), they have been identified as Goldstone modes \cite{WB04}, analogous to the familiar hydrodynamic modes and phonons. For the computation of the associated wave-vector dependent Lyapunov exponents governing 
the exponential time evolution
of the Lyapunov modes, random matrices \cite{eg00,TM02}, periodic-orbit expansion \cite{TDM02}, and,
most successfully, kinetic theory have been employed \cite{WB04,MM01,MM04}. 

\begin{figure}
\centering
{\includegraphics[width=8cm,angle=-90] {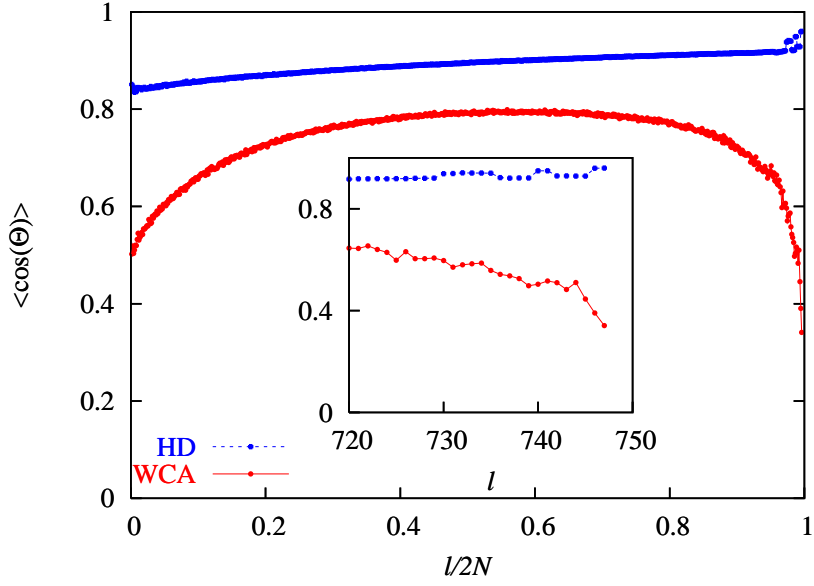}}\\
\caption{{\em Time-averaged} values for $\cos(\Theta_l)$, as defined in  Eq. (\ref{defcostheta}), for 
the indices $1 \leq l \leq 2N-3$ corresponding to the positive Lyapunov exponents of a spectrum. 
The systems consist of N = 375 particles at a density $\rho = 0.4$ in a rectangular periodic
box with aspect ratio $A = 0.6$;  Upper blue points: hard disks (HD); Lower red points: soft WCA disks.
In both cases, the temperature is unity.}
\label{costheta}
\end{figure}

For {\em hard-disk} systems the representation of the modes is simplified by the fact that, 
for large $N$, the position perturbations, $\{\delta q_i^{(l)}, i=1,\dots,N\}$, and momentum perturbations,
$\{\delta p_i^{(l)}, i=1,\dots,N\}$, may be viewed as two-dimensional vector fields $\varphi_q^{(l)}(x,y)$ and
$\varphi_p^{(l)}(x,y)$, respectively, which turn out to be nearly parallel (for $\lambda > 0$), or
anti-parallel  (for $\lambda < 0$) \cite{EFPZ04}. To illustrate this point, we plot in Fig. \ref{costheta}  
\begin{equation}
\left\langle\cos(\Theta_l)\right\rangle \equiv \left\langle\frac{\sum_{i=1}^N (\delta q_i^{(l)} 
\cdot \delta p_i^{(l)})} 
     { \sum_{i=1}^N (\delta q_i^{(l)})^2   \sum_{i=1}^N (\delta p_i^{(l)})^2 }\right\rangle ,
 \label{defcostheta}
 \end{equation}           
where $\Theta_l$ is the angle between the $2N$-di\-men\-sional vectors
comprising the  position and momentum perturbations of $\delta_l$, when they are viewed in the same 
$2N$-di\-men\-sional space. As always, $\langle \dots \rangle$ denotes a time average.
Only the indices $l < 2N-2$ corresponding to positive exponents are considered. $\Theta_l$
(upper blue curve) nearly vanishes  in the mode regime (close to the right-hand boundary
of that figure). For the negative exponents (not shown) the angle approaches $\pi$.
For large-enough $N$ and {\em small} positive $\lambda_l$, the individual particle
contributions behave as
$$
    \delta p_i^{(l)} = C_i^{(l)}(\Gamma) \delta q_i^{(l)},
$$    
where $C_i^{(l)}$ is a positive number, which is almost the same for all particles $i$.  
Once the $\delta q_i^{(l)}$ are known, the  $\delta p_i^{(l)}$ may be obtained from this relation. 
For a characterization of the modes, it suffices to consider only the position perturbations, 
which are interpreted as a vector field, $\varphi_q^{(l)}(x,y)$, 
over the simulation cell \cite{EFPZ04}.

For {\em soft-spheres}  the situation is more complicated. For low-density systems,
$\cos(\Theta_l)$ behaves as for hard disks, which is expected in view of the  similarity of the
Lyapunov spectra. For dense fluids, however, this quantity behaves qualitatively different and 
does not converge to unity but to zero in the interesting mode regime.  This is demonstrated in 
Fig. \ref{costheta} by the lower red curve, which is for $N=375$ soft WCA disks at the same  
density $\rho = 0.4$. The larger the collision frequency, the larger the deviations of 
$\langle\cos(\Theta_l)\rangle$ from 1 (for $\lambda_l>0$), or -1 (for $\lambda_l<0$) may become. 
This result indicates that
a representation of a Lyapunov mode in terms of a {\em single} (periodic) vector field of the position 
(or momentum) perturbations of the particles is possible only for low-density systems. For
larger densities and/or larger systems with many collisional events taking place at the same time,
one needs to simultaneously consider   all the perturbation components $\delta x_i, \delta y_i,
\delta p_{x,i}, \delta p_{y,i}$ of a particle.

\begin{figure}
\centering{\includegraphics[width=6cm,angle=-90] {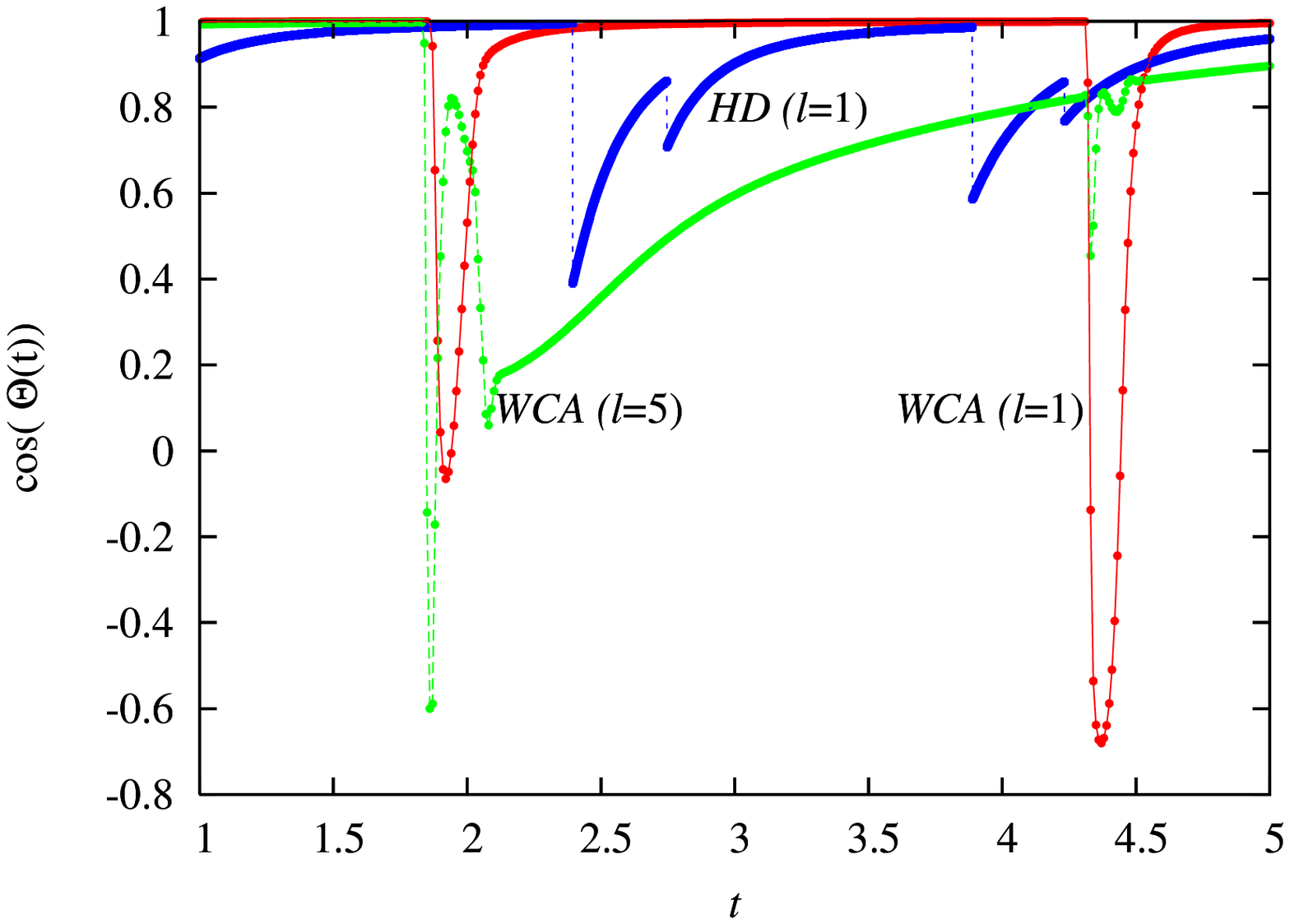}\\
                   \includegraphics[width=6cm,angle=-90] {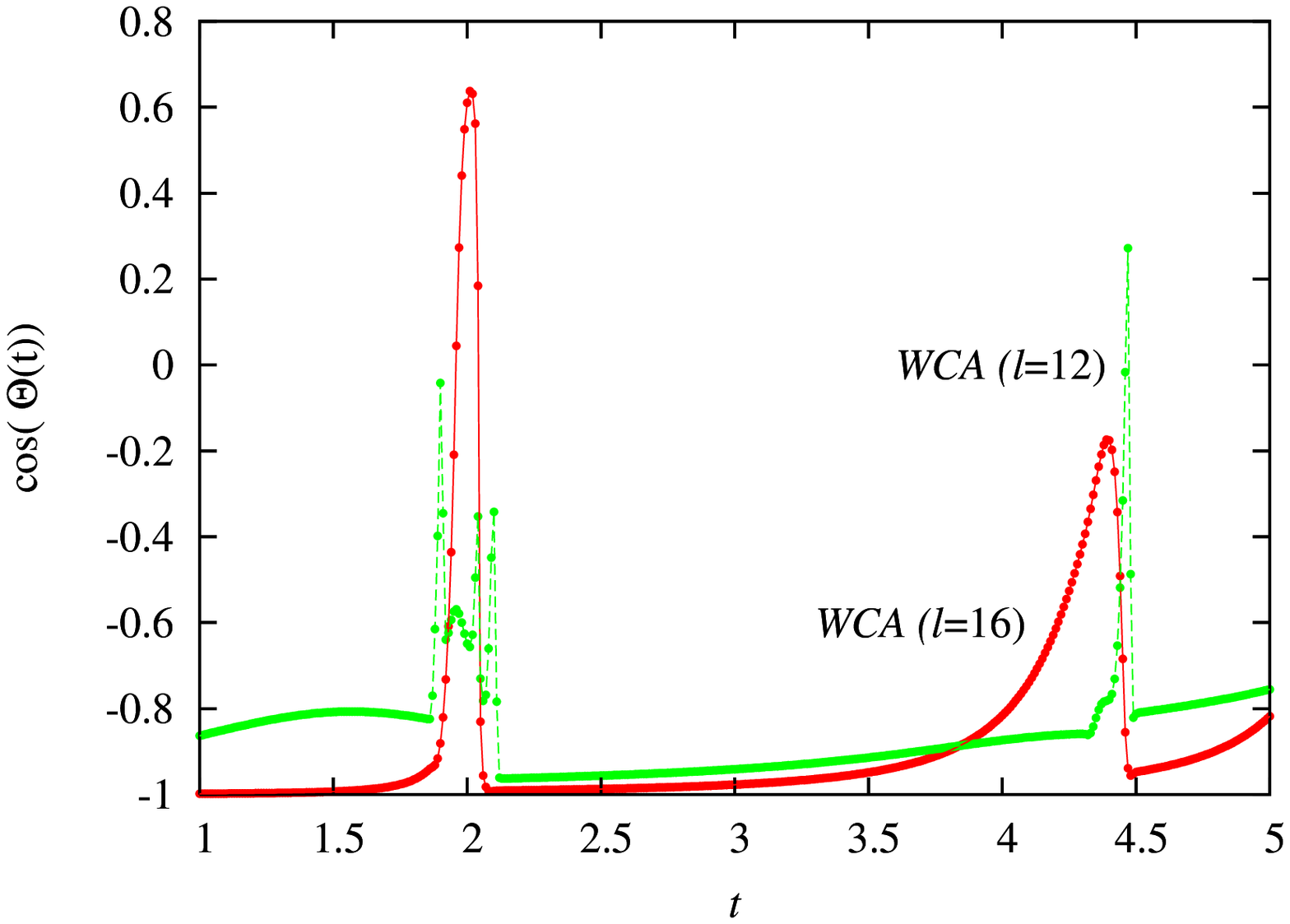}}
  \caption{Time dependence of  $\cos(\Theta_l)$ both in the time-forward (upper panel) and
  time-backward directions (lower panel) for four WCA particles in a square periodic box. 
  The density $\rho = 0.1$. The respective indices, $l$, are indicated by the labels. For
  comparison, an analogous hard-disk result for $l=1$ is also shown by the blue curve
  in the upper panel. }
  \label{costheta-t}
\end{figure}

To understand this unexpected behavior in more detail, we display in the top panel of 
Fig. \ref{costheta-t} the time dependences of $\cos(\Theta_l)$ belonging to the maximum $(l=1)$ 
and to the smallest-positive  $(l=2N-3 = 5)$ 
exponents of a four-particle system in a square periodic box with a density $\rho=0.1$.
During a streaming phase without collisions, the $2N$-dimensional vectors $\delta q^{(l)}$
and  $\delta p^{(l)}$ tend to become parallel as required by the linearized free-flight equations 
in tangent space. However,  this process is disrupted by a collision, which reduces 
(and in some cases  even reverses the sign of) $\cos(\Theta_l)$. In the following streaming
phase, i.e. forward in time, $ \Theta_l (t)$ relaxes towards zero. This suggests that the numerical
time evolution of   $\cos(\Theta_l)$ and, hence, of $\delta_l$ is not invariant with respect to time reversal. This is indeed
the case as is demonstrated in the lower panel of Fig. \ref{costheta-t}. To construct this
figure, the phase space trajectory was stored for another 10000 time units and, after a
time-reversal transformation, $\{q_i \to q_i, p_i \to -p_i; i=1,\dots,N\}$, was consecutively used 
- in reversed order - as the reference trajectory for the reversed tangent-space dynamics.
In the lower panel we show   $\cos(\Theta_{16})$ and
$\cos(\Theta_{12})$, which belong to the minimum and largest-negative exponents,
of the time-reversed dynamics, respectively, and which should be compared to  $\cos(\Theta_1)$ 
and $\cos(\Theta_5)$ for the forward evolution. Although the same collisions are involved,
the curves look totally different. This confirms previous results concerning the lack of
symmetry for the forward and backward time evolution of the Gram-Schmidt
orthonormalized tangent vectors $\delta_l$ for Lyapunov-unstable systems 
\cite{HHP01,HPH01}. Furthermore, we have numerically verified that 
$\cos(\Theta_l (t) ) = \cos(\Theta_{4N-l+1} (t) )$ is always obeyed, both forward and
backward in time.


\section{Fourier-transform analysis}
\label{FT}

Because of numerical fluctuations, a simple inspection of the vector field
$\varphi_q^{(l)}(x,y)$ very often is not sufficient to unambiguously 
determine the wave vector of a mode, particularly for the smaller than the maximum-possible
wave lengths.  Therefore, Fourier transformation methods
have been used \cite{Forster:2004}, where we regard $\varphi_a(x,y)$ 
as a spatial distribution
\begin{equation}
      \chi_a(q) = \sum_{i=1}^N a_i \delta(q - q_i).
      \label{distr}
\end{equation}   
The $a_i$ are identified, for example,  with $\delta x_i^{(l)}$  or with  \\
$\left(\gamma_i^{(l)}\right)^{1/2} \equiv  \left( (\delta x_i^{(l)})^2 + (\delta y_i^{(l)})^2 + (\delta p_{x,i}^{(l)})^2  + (\delta p_{y,i}^{(l)})^2\right)^{1/2}$.
In view of the periodicity of the box, the Fourier coefficients have wave numbers
given by Eq. (\ref{kvectors}). They are computed from 
\begin{equation}
     \chi_a({k}) \equiv \frac{1}{L_x L_y}\int d^2q e^{k \cdot q}    \chi_a(q) = 
     \frac{1}{L_x L_y}\sum_i a_i e^ {k \cdot q_i}.
\end{equation}
The power spectrum is defined by
\begin{equation}
  P_a({k}) =  \chi_a({k})  \chi_a({-k}).
 \end{equation}
For the $a_i$ identified above, the power spectra are denoted by  
 $ P_x(k)$ and $ P_{\gamma^{1/2}}(k)$, respectively.
We have also applied the algorithm for unequally-spaced points by Lomb  \cite{Lomb:1976,recipes}, 
suitably generalized to two-dimensional transforms.

\begin{figure}
\centering{\includegraphics[width=7cm,angle=-90]{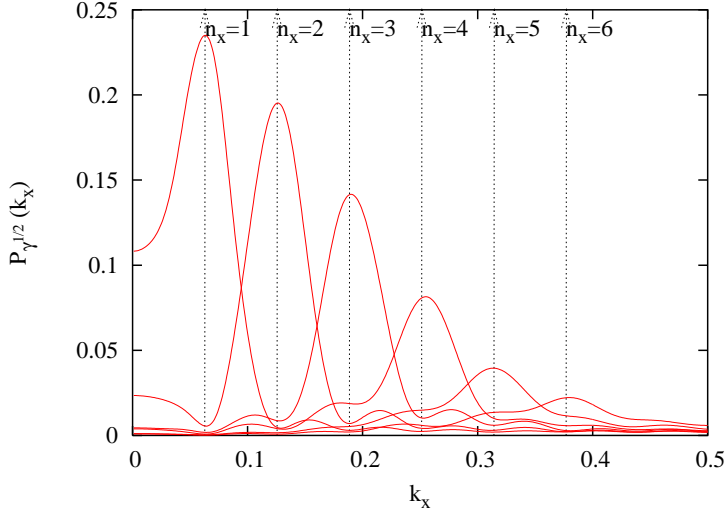}}
\vspace{3mm} \caption{Power spectra $ P_{qp}(k)$ for the transverse modes
T(1,0) to T(6,0) with  $l = 317$, 311, 305, 299, 293, and 287, for $N=160$ WCA particles
in an elongated  periodic box, $L_x = 100, L_y = 4$. The density  $\rho = 0.4.$
The Lyapunov spectrum for this system is shown in the lower panel of Fig. \ref{Fig_1}.}
\label{Lomb_0.4}
\end{figure}

\begin{figure}
\centering{\includegraphics[width=7cm,angle=-90]{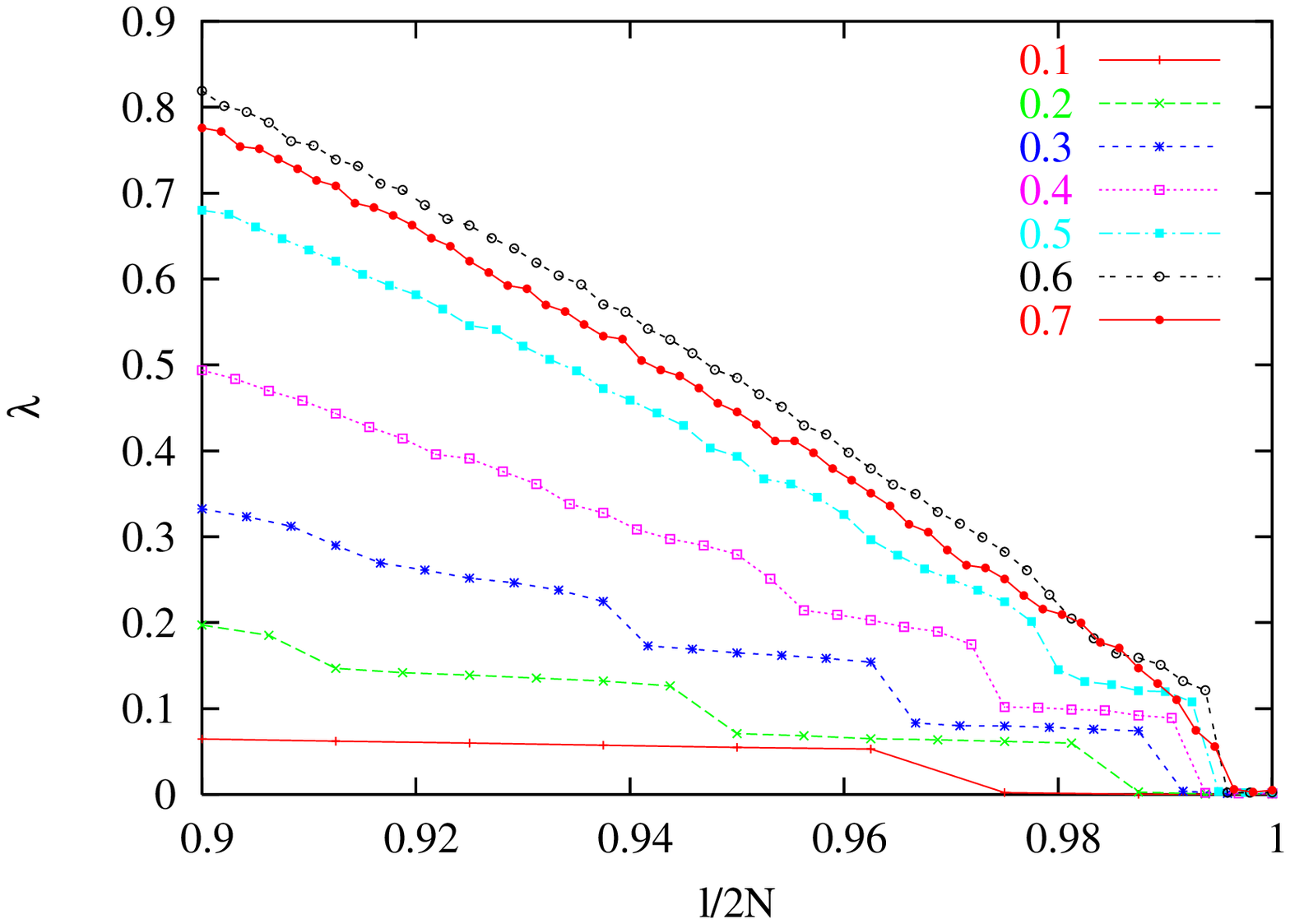}}\\
\centering{\includegraphics[width=7cm,angle=-90]{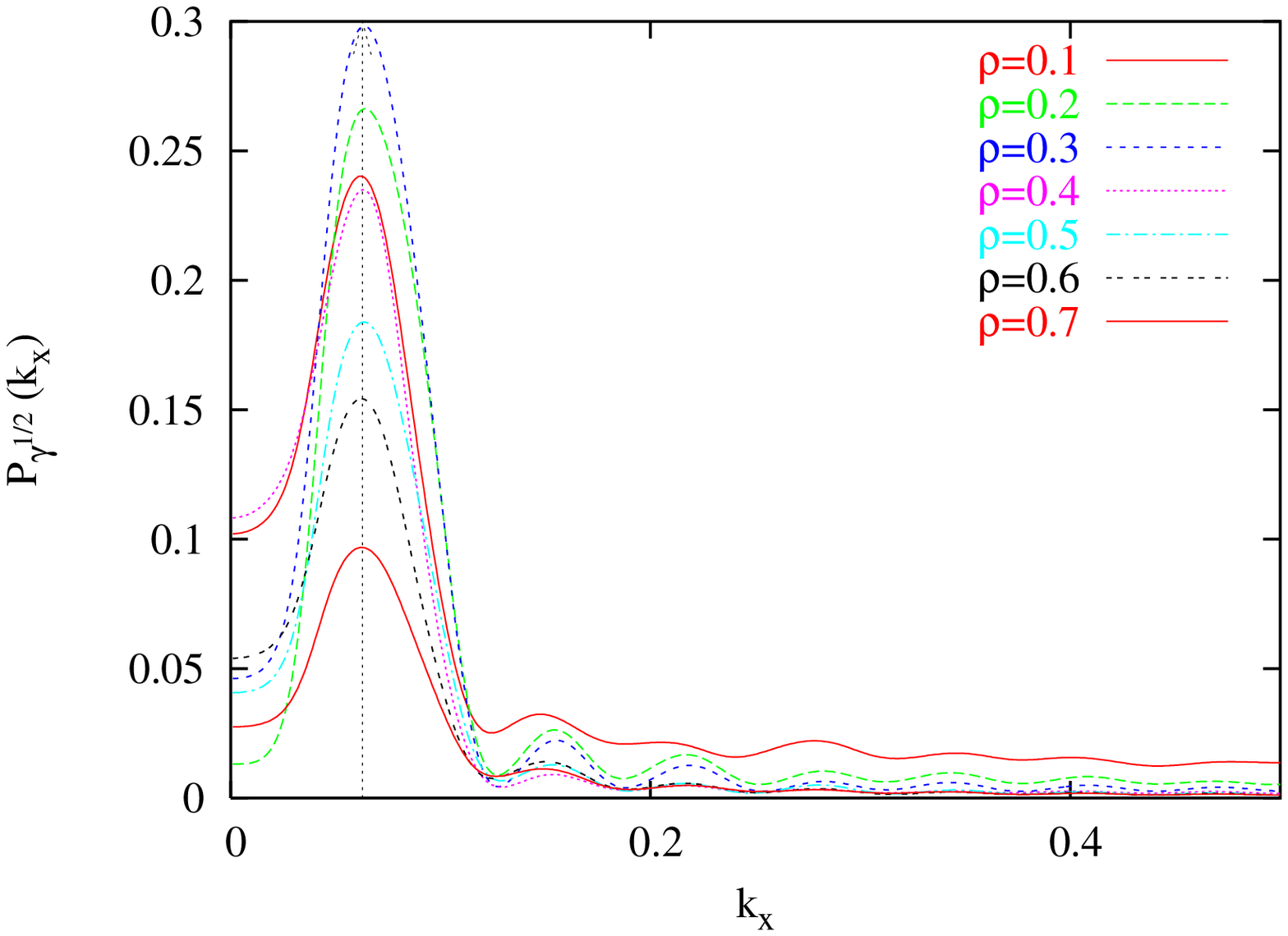}}
\vspace{3mm} \caption{WCA particles in a fixed periodic  box,  $L_x=100, L_y =4$, for various densities
as indicated by the labels. The temperature is unity. Upper panel:  Lyapunov spectra, plotted with
a reduced index $l/2N$ on the abscissa; 
Lower panel: power spectra  $ P_{\gamma^{1/2}}(k)$ for the modes corresponding to the 
smallest positive exponent, $\lambda_{l=2N-3}$, of each spectrum.} 
\label{full_lomb}
\end{figure}

As a first  example we show in Figure \ref{Lomb_0.4} the power spectra $ P_{\gamma^{1/2}}(k)$ for
successive transverse modes, indexed by  $l = 317$, 311, 305, 299,  293, and 287, 
for the WCA-system described in the lower panel of Fig. \ref{Fig_1}. They 
correspond to the successive wave numbers $k_{1,0}$ to $k_{6,0}$ as 
predicted by Eq. (\ref{kvectors}). In a second and  less-trivial example, we show in
Fig. \ref{full_lomb} the Lyapunov spectra (left panel) of WCA particles in a
fixed elongated box, $L_x = 100, L_y=4$, for various densities varying from 
$\rho = 0.1$ to  $\rho = 0.7$ as indicated by the labels. 
In the right panel the power spectra  $ P_{\gamma^{1/2}}(k)$ for the modes corresponding to the 
smallest positive exponent, $\lambda_{l=2N-3}$, of each spectrum are shown. Since $L_x$
is the same in all cases, all power spectra have a peak at the allowed wave number
$k_{1,0} = 0.063$. This peak is also well resolved for large densities, for which the
step structure in the spectrum is blurred.


\section{Concluding remarks}
\label{remarks}

\begin{figure}
\centering{\includegraphics[width=7cm,angle=-90]{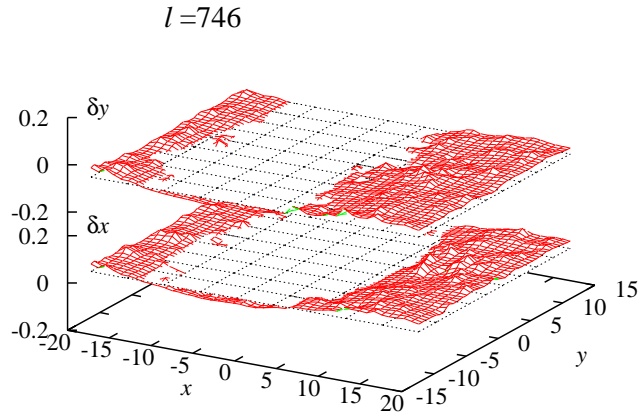}}
\vspace{3mm}
\caption{Representation of a Lyapunov mode as periodic spatial patterns of the 
position perturbations $\delta x_i$ and $\delta y_i$ of the particles in the periodic box.
The system consists of 375 WCA disks at a density $\rho = 0.4$ in a rectangular box with an
aspect ratio $A = 0. 6$. The Lyapunov spectrum for this system is given by the top-most 
curve of Fig. \ref{Fig_2}. The mode $l=746$ is shown.}
\label{fig_mode}
\end{figure}

        In the foregoing sections we have demonstrated the existence of Lyapunov modes
in soft-disk fluids with the help of various indicators such as the localization measure in Sec.
\ref{sec_loc}, and the Fourier analysis in the previous section. However, the classification 
and characterization of the modes is more complicated than for hard disks. This is demonstrated
in Fig. \ref{fig_mode}, where the position perturbations $\delta x_i^{746}$ (bottom) and 
$\delta y_i^{746}$ (top) are plotted at the particle positions in the simulation cell for the 
375-particle system with a density $\rho = 0.4$ familiar from Fig. \ref{Fig_2}. Naively, the upper surface 
would have to be classified as a transverse mode with the perturbation component
$\delta y$ perpendicular to a wave vector $k$ parallel to $x$, and the lower surface as a longitudinal 
mode with $\delta x$ parallel to $k$. Since the momentum perturbations are not nearly 
as parallel to the position perturbations (see Fig. \ref{costheta}) as is (approximately) the case for hard disks, it is not sufficient to represent a mode by  a single two-dimensional vector field  such as $\varphi_q^{(l)}(x,y)$ or $\varphi_p^{(l)}(x,y)$.  Position and momentum perturbations need to be
considered simultaneously for a characterization of the modes. The disappearance of the
step structure in the Lyapunov spectrum and, hence, of the degeneracy for larger densities points to a complicated tangent-space dynamics we have not yet been able to unravel. We do not
know of any theory which accounts for all of the numerical results presented in this paper.

\section*{Acknowledgments}
We gratefully acknowledge fruitful discussions with C. Dellago, J.-P. Eckmann, R. Hirschl, 
Wm. G. Hoover, H. van Beijeren, and E. Zabey, and with participants of two recent  workshop,
one at CECAM in Lyon, in July 2004, and one at the Erwin Schr\"odinger Institute (ESI) in Vienna, 
in August 2004, where part of our results were presented. The workshop at ESI was co-sponsored by the European Science Foundation within the framework of its STOCHDYN program. Our  work was supported by the Austrian Fonds zur F\"{o}rderung der Wissenschaftlichen Forschung, grant $P15348-PHY$. We also thank the Computer Center of the University of Vienna for  a generous allocation of 
computer ressources at the computer cluster ``Schr\"{o}dinger II''.

\bibliographystyle{unsrt}

\end{document}